\documentclass[lettersize,journal]{IEEEtran}
\usepackage{amsmath,amsfonts}
\usepackage[ruled,norelsize,linesnumbered]{algorithm2e}
\usepackage{float}
\makeatletter
\newcommand{\removelatexerror}{\let\@latex@error\@gobble}
\makeatother
\usepackage{array}
\usepackage[caption=false,font=footnotesize,labelfont=sf,textfont=sf]{subfig}
\usepackage{textcomp}
\usepackage{stfloats}
\usepackage{url}
\usepackage{verbatim}
\usepackage{cite}
\usepackage{graphicx}
\usepackage{multirow}
\usepackage{color}
\usepackage{booktabs}

\def\BibTeX{{\rm B\kern-.05em{\sc i\kern-.025em b}\kern-.08em
    T\kern-.1667em\lower.7ex\hbox{E}\kern-.125emX}}
\usepackage{balance}
\begin{document}

\title{Few-Shot Recognition and Classification Framework for Jamming Signal: A CGAN-Based Fusion CNN Approach}


\author{Xuhui Ding,~\IEEEmembership{Member,~IEEE}, Yue Zhang, Gaoyang Li, Xiaozheng Gao,~\IEEEmembership{Member,~IEEE}, Neng Ye,~\IEEEmembership{Member,~IEEE}, Dusit Niyato,~\IEEEmembership{Fellow,~IEEE}, and Kai Yang,~\IEEEmembership{Member,~IEEE}
	\thanks{
		
		Xuhui Ding and Neng Ye are with the School of Cyberspace Science and Technology, Beijing Institute of Technology, Beijing 100081, China (e-mail: dingxuhui@bit.edu.cn; ianye@bit.edu.cn).	
		
		Yue Zhang, Xiaozheng Gao, and Kai Yang are with the School of Information and Electronics, Beijing Institute of Technology, Beijing 100081, China (e-mail: yuez@bit.edu.cn; gaoxiaozheng@bit.edu.cn; yangkai@ieee.org).			
		
		Gaoyang Li is with School of Biological Science and Medical Engineering, Beihang University, Beijing 100191, China (e-mail: a798876105@gmail.com).

		Dusit Niyato is with the School of Computer Science and Engineering, Nanyang Technological University, Singapore 639798 (e-mail: dniyato@ntu.edu.sg).

}
}

\maketitle

{
\begin{abstract}
	 Subject to intricate environmental variables, the precise classification of jamming signals holds paramount significance in the effective implementation of anti-jamming strategies within communication systems. 
	 In light of this imperative, we propose an innovative fusion algorithm based on conditional generative adversarial network (CGAN) and convolutional neural network (CNN), which aims to deal with the difficulty in applying deep learning (DL) algorithms due to the instantaneous nature of jamming signals in practical communication systems. Compared with previous methods, our algorithm embeds jamming category labels to constrain the range of generated signals in the frequency domain by using the CGAN model, which simultaneously captures potential label information while learning the distribution of signal data thus achieves an 8\% improvement in accuracy even when working with a few-sample dataset. Real-world satellite communication scenarios are simulated by adopting hardware platform, and we validate our algorithm by using the resulting time-domain waveform data. The experimental results indicate that our algorithm still performs extremely well, which demonstrates significant potential for practical application in real-world communication scenarios.
	 
\end{abstract}
}
\begin{IEEEkeywords}
communication anti-jamming, jamming signal classification, convolution neural network, generative adversarial network, neural network algorithm.
\end{IEEEkeywords}

\section{Introduction}
\IEEEPARstart{T}{he} reliable transmission of signals through satellite communication links is a fundamental requirement to ensure the regular functioning of human activities on earth\cite{9444629,10210347}. The normal operation and maintenance of satellite communication systems is a fundamental prerequisite for providing wide coverage, secure, high-speed data transmission, and diverse information services. The Global Navigation Satellite System (GNSS), composed of four navigation systems, namely Global Positioning System (GPS), Beidou System (BDS), Galileo, and GLONASS, offers global coverage navigation services with high-precision positioning capabilities, which find broad applications in civilian and military domains. The emerging star link system provides effective data transmission and high-speed internet services by launching a large number of low-orbit satellites to establish a global satellite communication network. However, the satellite communication system confronts complex environmental factors such as solar flares, the Earth's magnetic field, and the atmosphere, which lead to signal instability and inevitable signal attenuation\cite{9789278,10296886}. Electromagnetic and radio-frequency jamming further weaken the quality of effective signals during transmission. Therefore, the adoption of anti-jamming measures is crucial to ensure the normal information transmission of satellite communication systems. 

Recent studies have explored new techniques in the wireless communication field to enhance the robustness of wireless transmission. Minimizing external jamming during channel transmission is essential to ensure accurate signal exchange between communication parties. Electromagnetic noise and human jamming can result in a higher bit error rate (BER) during signal demodulation at the receiving end, leading to ineffective communication scenarios. Jamming and anti-jamming technologies have been the subject of extensive research in communication engineering. The standard approach involves implementing suitable anti-jamming measures to counter specific jamming modes. Wang et al. \cite{Wang2013JammingEO} quantitatively assessed the impact of jamming on navigation satellite links. Li et al. \cite{Li2023InfluenceOS} discussed the effect of swept jamming on time-domain adaptive anti-jamming and proposed an adaptive filtering timing reset method to suppress interfering signals. Liu et al. \cite{Liu2022SidelobesSF} proposed a correlation function side-flap suppression method to improve the anti-jamming and capture performance of satellite navigation receivers. In recent decades, scholars have widely proposed intelligent anti-jamming strategies, such as reinforcement learning, to optimize anti-jamming strategies, making them more flexible and intelligent. For instance, Arif et al. \cite{Arif2021ASR} employed reinforcement learning to empower satellite communication systems with intelligent resource management and anti-jamming capabilities. Proper jamming type identification is a prerequisite for implementing necessary anti-jamming measures. 

A strategic system aimed at anti-jamming typically comprises three primary stages: jamming detection, jamming identification, and anti-jamming strategy optimization. The correct identification of jamming type directly affects the accurate implementation of anti-jamming methods. Traditional classification algorithms, such as the smoothed pseudo-Wigner–Ville distribution (SPWVD)\cite{5424070}, support vector machine (SVM)\cite{Cortes1995SupportVectorN}, decision tree (DT)\cite{Salzberg1994C45PF}, and back propagation neural network (BP)\cite{Rumelhart1986LearningRB}, have strengths and weaknesses regarding classification performance, which can vary on different data samples. The emergence of deep learning (DL) algorithms has recently significantly impacted recognition and classification tasks, {and they have been already applied in wireless communication scenarios\cite{HuynhThe2021AutomaticMC,9622204,Bai2022ElectromagneticMS,10001441,10264147,9697978,10003212}}.
\par {However, DL algorithms are data-driven models that depend on high-capacity and high-precision datasets. The datasets used in most DL-based jamming identification methods are capable of supporting high accuracy in identification. Nevertheless, in real-world scenarios, obtaining a large amount of labeled data in a short period may not be available, which leads to a sharp degradation of the performance of DL-based jamming signal identification methods.}

To the best of our knowledge, this is the first work to achieve {jamming signal identification on a few-shot dataset in realistic scenarios. The main contribution of this study is summarized as follows}.

\begin{itemize}

	\item{A one-dimensional (1D) preprocessing method with lower time complexity is employed to enhance the real-time performance of the algorithm for practical application scenarios.}
	\item {We integrate the one-dimensional conditional generative adversarial network (1D-CGAN) and one-dimensional conditional generative adversarial network (1D-CNN) as a fusion algorithm, which exhibits remarkable performance to recognize jamming patterns on few-shot datasets characterized by low precision data and limited capacity.}
	\item {We construct a hardware platform to acquire real-world datasets from simulated communication links. Comparative experimental results demonstrate that our proposed algorithm can achieve higher-accuracy recognition of jamming patterns even when the analog-to-digital converter (ADC) acquires as few as 30 samples under each class.}
\end{itemize}
\par The remainder of this paper is structured as follows. Section II primarily presents the research progress on jamming recognition technology and generative adversarial network (GAN). Section III provides the description and mathematical models of common jamming signals. The method of our proposed model is described in Section IV. Section V provides the details of training objectives and methods. The experimental parameters and simulation results are presented in Section VI and Section VII, respectively. Finally, we conclude our work in Section VIII.
\section{Recent Relevant Work}
\par {In this section, we first introduce the target problem to be addressed, which encompasses the relevant research and methods in jamming identification. Next, we delve into recent GAN-related research that is pertinent to the methodology employed in this paper.}
\subsection{Jamming Signal Identification Methods}
The fundamental principle of jamming signal classification is to extract discriminative features from signal sequences with diverse characteristics. Traditional approaches include SPWVD, BP, SVM, DT, and generalized likelihood ratio test (GLRT)\cite{7762200}, among others. For instance, Greco et al. \cite{Greco2008RadarDA} evaluated the performance of the Adaptive Coherence Estimator (ACE) and GLRT in the context of radar signal jamming detection and classification. Moreover, some innovative feature extraction algorithms based on the frequency domain have been introduced \cite{Xiaoyan2012PatternRM,Yongcai2014AFT,Hao2019RecognitionMO}. Nevertheless, conventional classification algorithms are characterized by high complexity and necessitate prior knowledge of signal processing and data structure. In this regard, {researchers have proposed employing DL algorithms to detect and identify jamming types due to their lower complexity, lack of prior knowledge requirements, and potential for achieving higher performance. Swinney et al. \cite{Swinney2021GNSSJC} achieved detecting and classifying RF jamming signals from GNSS by utilizing the transfer learning (TL) algorithm to extract features from image information.} Specifically, for signal sequences, {1D-CNN} can be utilized for feature extraction, which is simpler and does not require a deep theoretical understanding of complex signals. For instance, Sun et al. \cite{Sun2018RadarEC} proposed the application of 1D-CNN to train a large dataset of radar transmitter data and subsequently classify high-dimensional sequences. Junfei et al. \cite{Junfei2018BarrageJD} introduced a novel method for SAR jamming detection and classification that relies on CNN\cite{LeCun1998GradientbasedLA}. Furthermore, Xu et al. \cite{Xu2020AJR} developed two jamming classifiers based on DT and deep neural network (DNN), respectively, and compared their jamming detection performance in the context of satellite navigation systems, where the comparison results showed that the DNN algorithm outperformed the DT algorithm significantly.
 Zhang et al. \cite{Zhang2021FastCC} explored a fast complex-valued CNN (F-CV-CNN) algorithm based on pruning, which simultaneously improves the accuracy and speed of jamming signals. {Some researchers also use traditional GAN structures for jamming detection or anti-jamming spectrum access \cite{9514852,Fan2020DeceptiveJT,LuoLL0Y22,Huang2023SpaceBasedES}.}
 
 {In parallel, to tackle the difficulty of attaining enhanced performance using DL algorithms in realistic scenarios, several researchers have put forth innovative solutions. For instance, Hou et al. \cite{Hou2022RadarJammingCI} have introduced the TL algorithm as a few-shot leaning approach to precisely classify radar jamming signals, even when working with limited training data. While the algorithm demonstrates remarkable accuracy, it is important to note that the authors have acknowledged certain limitations, including its applicability solely to simulated data and specific data volume prerequisites.
   Another noteworthy contribution in this domain is the work by Shao et al. \cite{Shao2020ConvolutionalNN}, which presents a CNN-based Siamese network for the classification of radar jamming signals. This innovative approach has shown promise in achieving satisfactory classification results, even with a so limited training dataset. Nonetheless, the ongoing challenge lies in devising a novel methodology that strikes a balance between classification accuracy and the minimal number of training instances. To be more precise, there is an unmet need to enhance classification accuracy while working within the constraints of a limited training dataset.}
  \begin{table}[]
 	\caption{Main Notations Used in This Paper}
 	\centering
 	\begin{tabular}{ll}
 		\toprule[1.2pt]
 		\textbf{Notation}&\textbf{Description}\\
 		\midrule
 		$x$&The preprocessed 1D vector\\
 		\midrule
 		$y$&The embedded label information\\
 		\midrule
 		$z$&The random sampling noise\\
 		\midrule
 		$G(z,y)$&The output of the generator\\
 		\midrule
 		$D(G(z,y),y)$&The output of the discriminator when receiving\\
 		&$G(z,y)$ and $y$\\
 		\midrule
 		$D(x,y)$&The output of the discriminator when receiving\\
 		&$x$ and $y$\\
 		\midrule
 		$C_i$&The prediction results belonging to category $i$\\
 		\midrule
 		$N$&The batch size during training period\\

 		\midrule
 		$L_2$&The length of $x$ and $G(z,y)$\\
 		\midrule
 		$L_1$&The length of $z$\\
 		\midrule
 		$P_J$&The Power of jamming signal\\
 		\midrule
 		$f_J$&The center frequency of jamming signal\\
 		\midrule
 		$\theta_J$&The initial phase of jamming signal\\
 		\midrule
 		$U_0$&The carrier amplitude of jamming signal\\
 		\midrule
 		$v_0^i,v_1^i$&The $i$-th element of a vector consisting \\
 		&entirely of zeros or ones\\
 		\toprule[1.2pt]
 	\end{tabular}	
 \end{table} 
  {
  Upon the above observations, the difference in preprocessing methods will directly affect the classification results of the algorithm. Some approaches use time-frequency images (TFIs) or frequency-domain images (FIs) for classification. However, the limitations of this method stem from its high complexity and a need for prior knowledge of signal processing.  Hence, we avoid the tedious image-processing approach and propose to use 1D data that only requires the fast Fourier transform (FFT) results instead of two-dimensional (2D) image data preprocessing to address these limitations.}

\subsection{Development and Research of GAN}
Recently, GAN has successfully generated images that are almost indistinguishable from actual samples and has had a profound impact on natural language generation. {Mirza et al. \cite{2014Conditional} introduced conditional information and proposed a CGAN structure that can be used to control the direction of image generation.} While images are typically fed into the GAN network as 2D data, signal sequences are simpler and are typically represented as 1D data. Thus, many improved GAN models have been proposed to process 1D data. Zec et al. \cite{zec2019recurrent} implemented the Recurrent Conditional Generation Adversarial Network (RC-GAN), which utilizes a recurrent neural network consisting of long short-time memory (LSTM)\cite{Hochreiter1997LongSM} in both the generator and discriminator to generate time series sensor errors exhibiting long-term time correlation. Mogren et al. \cite{Mogren2016CRNNGANCR} proposed a generative adversarial model based on GAN that operates on continuous sequential data to generate classical music. Esteban et al. \cite{Esteban2017RealvaluedT} proposed a Recurrent GAN (RGAN) and a Recurrent Conditional GAN (RCGAN) to generate real multidimensional time series and focused on their application to medical data. Donahue et al. \cite{Donahue2018SynthesizingAW} introduced WaveGAN, the first attempt to apply GAN to original audio synthesis in an unsupervised environment. Li et al. \cite{Li2022TTSGANAT} introduced TTS-GAN, a transformer-based GAN capable of generating 1D data of finite length, which provides a novel approach to generating jamming signal samples for data preprocessing. Additionally, there are examples of using GAN to denoise ECG signals, such as NR-GAN proposed by Sumiya et al. \cite{Sumiya2019NRGANNR} with significant effects. {Considering the manifold advantages of GAN, we introduce a novel algorithmic model based on CGAN that exhibits the capability to generate identically distributed 1D data sequences of variable length by learning from 1D training datasets. Consequently, the model manifests a high degree of congruity with the statistical attributes of signals. Notably, the proposed model achieves superior performance by utilizing only a limited number of samples as input to predict the types of jamming signals in communication systems accurately. In Section IV, we elaborate on the proposed methodology and its applications. The commonly used notations are displayed in Table I.}
\section{System Model}

\par {In this section, we provide mathematical expressions for different jamming signals and visualize their corresponding time and frequency domain characteristics.}

A comprehensive discussion is carried out including narrowband noise-, band noise-, analog modulation-, and periodic pulse noise jamming, etc. Obtained simulation results are utilized as training samples and input into the algorithmic model, and the simulation process follows the same approach as in the work\cite{Xu2020AJR}. The mathematical expression of Continuous Wave Jamming (CWJ) is as follows:
\begin{equation}
	j(t)=\sqrt{P_{J}} \exp \left[j\left(2 \pi f_{J} t+\theta_{J}\right)\right],\label{eq1}
\end{equation}
where $P_{J}$ is the jamming power and $f_{J}$ is the jamming center frequency, $\theta_{J}$ is the random initial phase which obeys the uniform distribution on $\left[0,2 \pi\right)$. In this paper, we control the difference in amplitude and frequency to divide the CWJ pattern into two different categories {called $CWJ_{A}$ and $CWJ_{W}$ }to detect the classification effect of the algorithm. 
\begin{figure}[!t]
	\centering
	\includegraphics[width=3.4in]{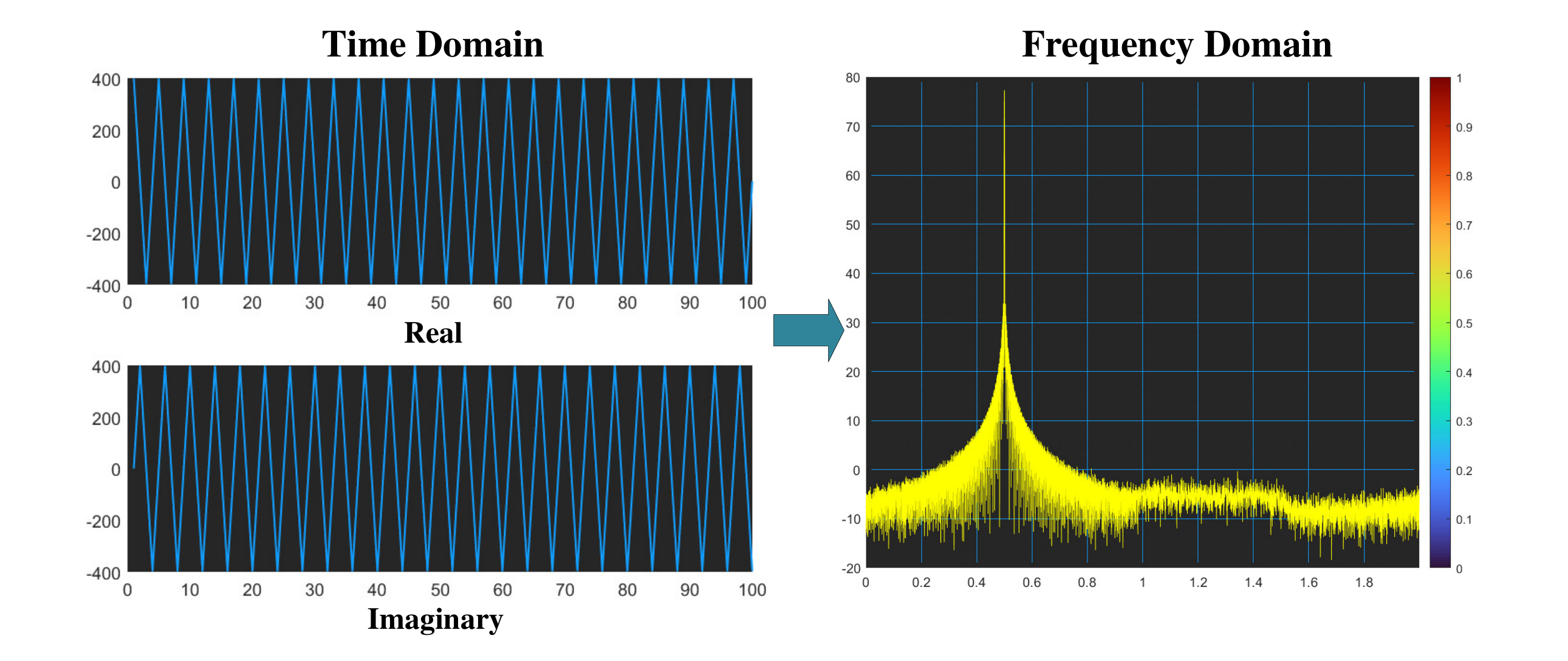}
	\caption{Time and frequency domain images of the CWJ.}
	\label{fig1}
\end{figure}

\begin{figure}[!t]
	\centering
	\includegraphics[width=3.4in]{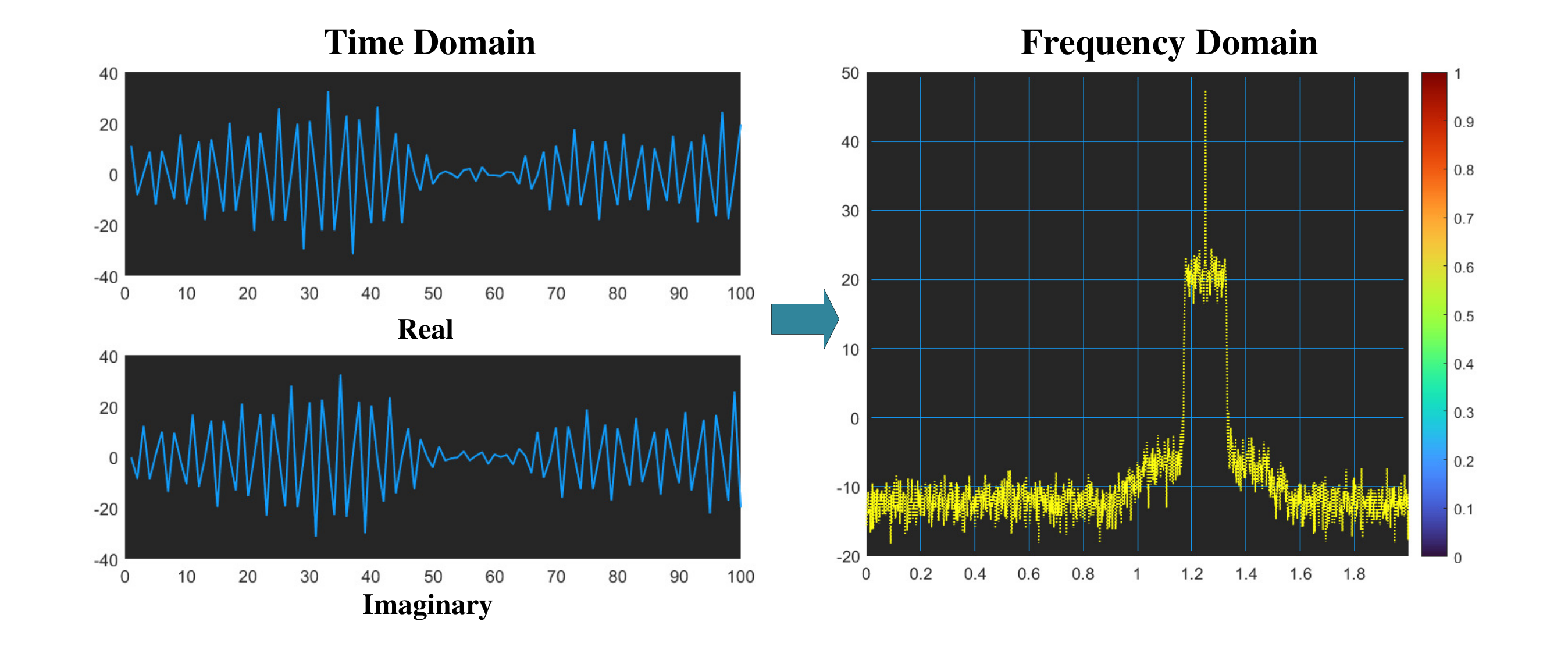}
	\caption{Time and frequency domain images of the NAMJ.}
	\label{fig3}
\end{figure}
Traditionally, Amplitude Modulation Jamming (AMJ) is produced through modulation of the carrier's amplitude by a cosine signal, which is mathematically expressed as:
\begin{equation}
	j(t)=\sqrt{\frac{P_J}{U_0+\beta_{A M}^2}}\left(U_0+m(t)\right)\exp\left[j\left(2\pi f_J t+\theta_J\right)\right],\\
\end{equation}
\begin{equation}
	m(t)=\beta_{A M} \cos \left(2 \pi f_m t+\theta_m\right), \\
\end{equation}
where $m(t)$ is the modulation signal, {$\beta_{A M}$ represents the peak-to-peak value of the modulated signal, which is called the modulation index}, $\theta_m$ is the random initial phase of the modulation signal, which obeys the uniform distribution on $\left[0,2 \pi\right)$, and ${U_0}$ stands for the carrier amplitude. 
\begin{figure}[!t]
	\centering
	\includegraphics[width=3.4in]{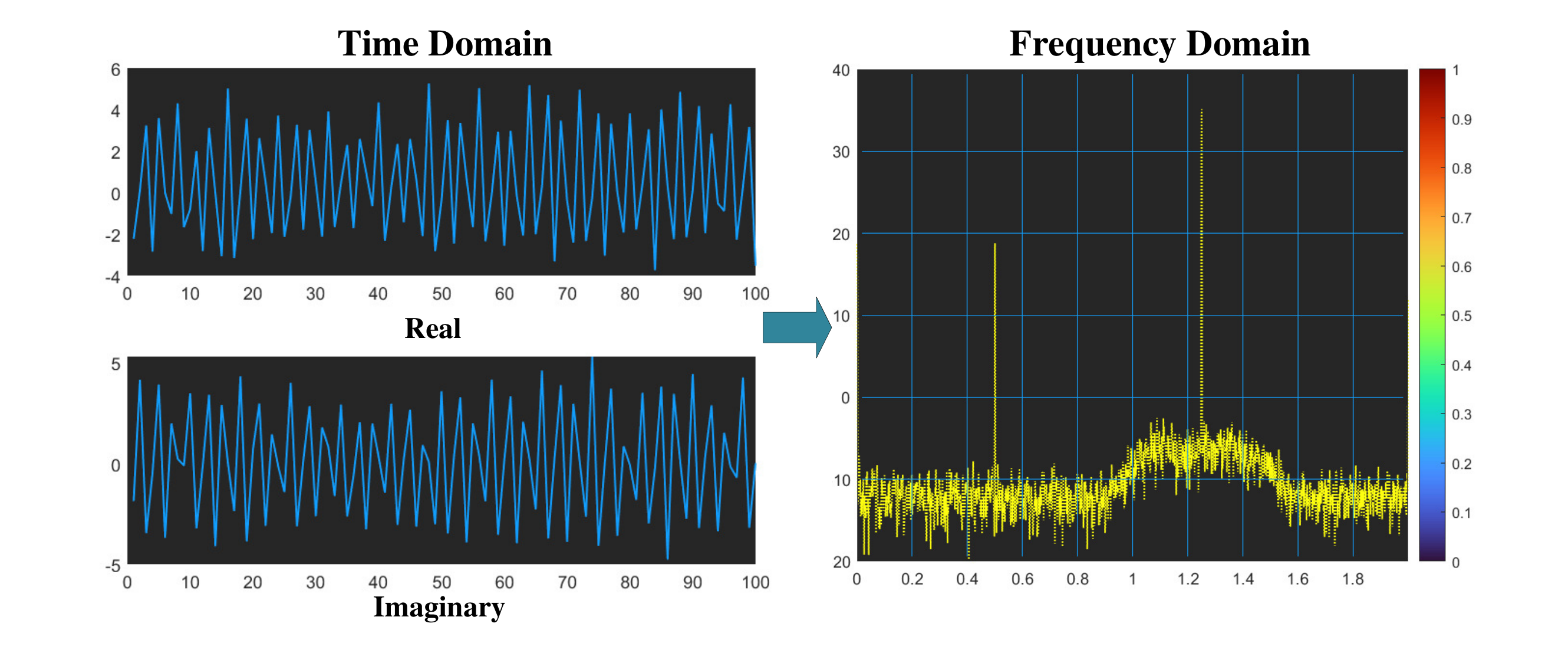}
	\caption{Time and frequency domain images of the AMJ.}
	\label{fig2}
\end{figure}
Noise Amplitude Modulation Jamming (NAMJ){\cite{9416448}} is obtained by amplitude modulation of the carrier by baseband noise:

\begin{equation}
	j(t)=\left[ {U_0+U_n(t)}\right]\exp \left[j\left(2 \pi f_J t +\theta_J\right)\right],
\end{equation}
where $U_n(t)$ stands for the baseband noise.

The Narrow Band Noise Jamming (NBNJ) is obtained by the frequency shift of narrowband noise:
\begin{equation}
	j(t)=U_n(t)\exp \left[j\left(2 \pi f_J t +\theta_J\right)\right].
\end{equation}
\begin{figure}[!t]
	\centering
	\includegraphics[width=3.4in]{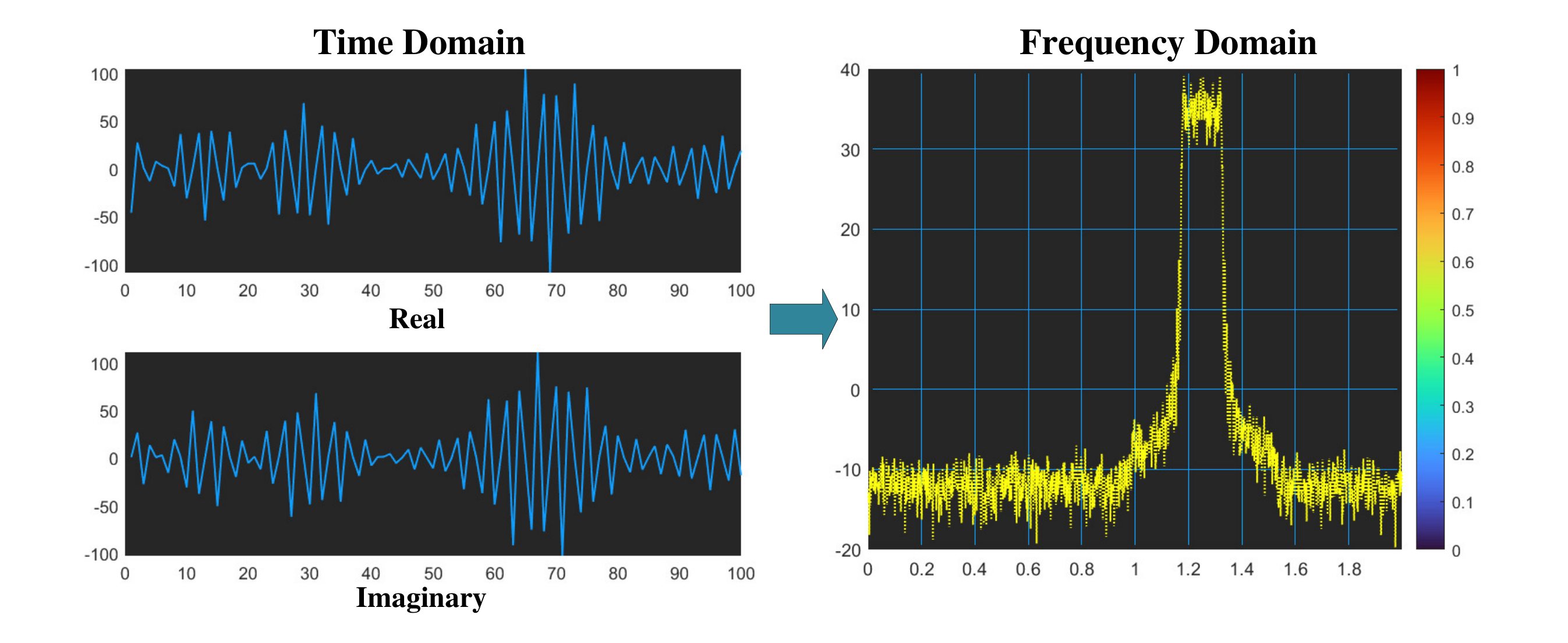}
	\caption{Time and frequency domain images of the NBNJ.}
	\label{fig4}
\end{figure}
\par Multitone Jamming (MTJ) is a type of Random Comb Jamming (RCJ) \cite{Xu2020AJR}, which is generated by superimposing several single tones with different frequencies. The mathematical expression of MTJ can be represented as follows:
\begin{equation}
	j(t)=\sum_{i=1}^{N_T} \sqrt{P_T(i)} \exp \left[j\left(2 \pi f_T(i) t+\theta_i\right)\right],
\end{equation}
{where $N_T$ represents the number of different tones, which serves as a variable simulation parameter to generate different datasets. $P_T(i)$ is the power of the $i$-th tone's frequency, and $f_T(i)$ is the center frequency of the $i$-th tone. $\theta_i$ is the initial phase of the $i$-th tone}, $\theta_i \sim \left[ 0, 2 \pi\right)$.

\begin{figure}[!t]
	\centering
	\includegraphics[width=3.4in]{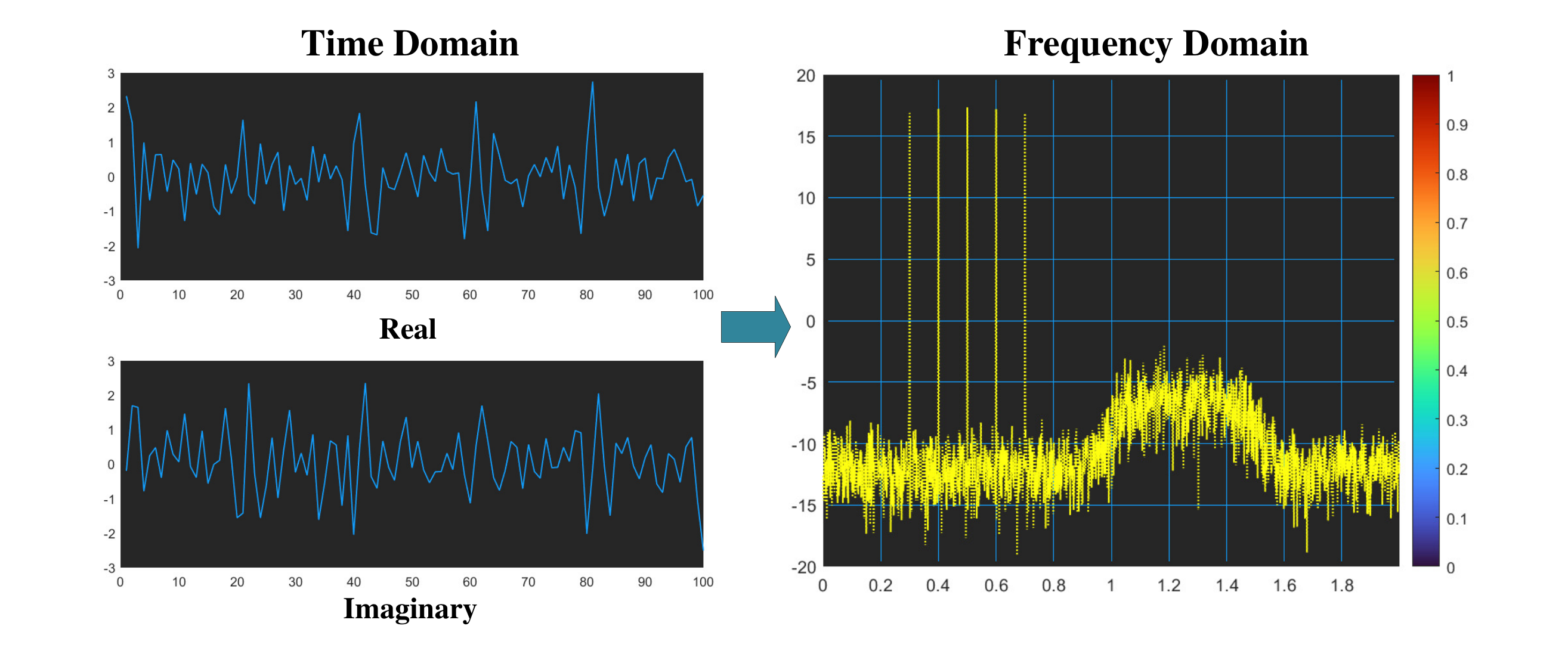}
	\caption{Time and frequency domain images of the MTJ.}
	\label{fig5}
\end{figure}
Linear Frequency Modulation Jamming (LFMJ) is a type of jamming in which the carrier frequency varies linearly with time, resulting in what is also known as linear sweep jamming. The mathematical expression for LFMJ is as follows:
\begin{equation}
	j(t)=\sqrt{P_J} \exp \left\{j\left[2 \pi\left(f_L+\frac{f_H-f_L}{2 \times T_{s w}} t\right) t+\theta_J\right]\right\},
\end{equation}
where $f_L$ and $f_H$ represent the initial frequency and cut-off frequency of the jamming frequency band, $T_{sw}$ is the sweep period, the sweep bandwidth $W_{sw}=f_H-f_L$, and the frequency modulation rate is $\frac{f_H-f_L}{2 T_{sw}}$.

\begin{figure}[!t]
	\centering
	\includegraphics[width=3.4in]{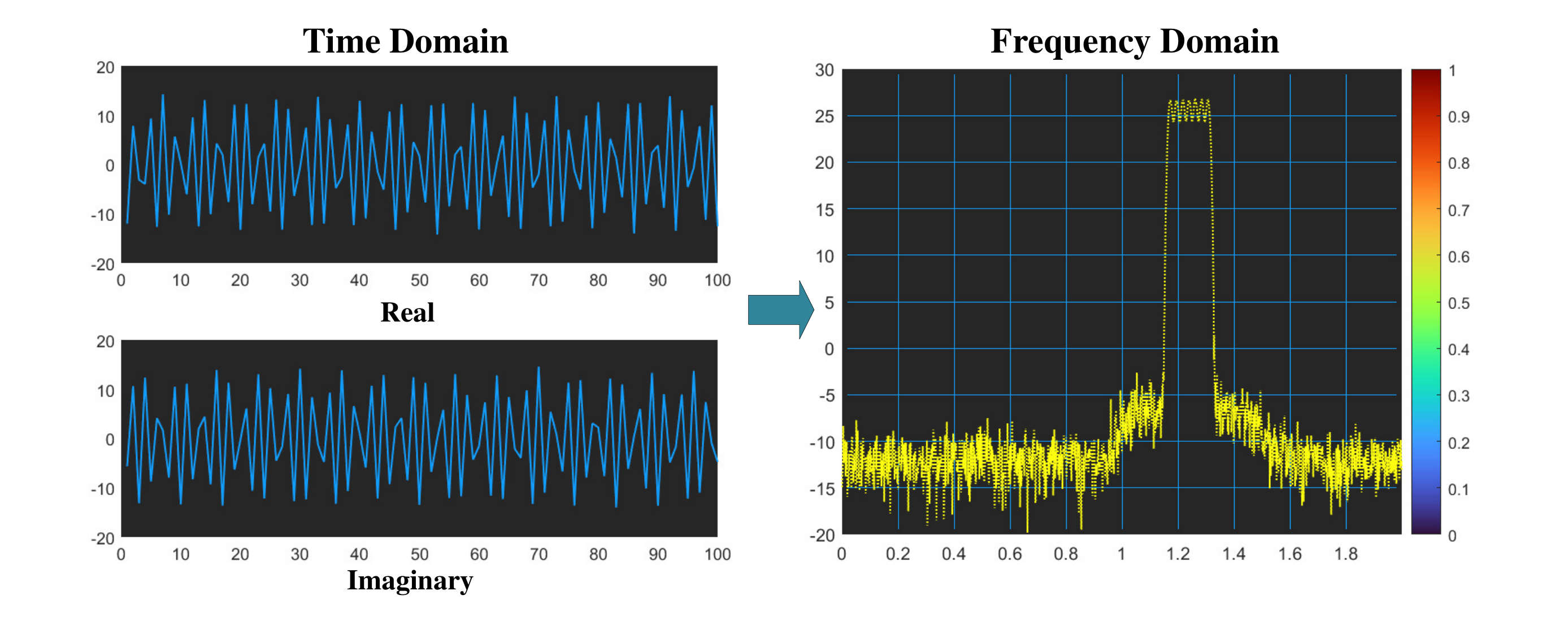}
	\caption{Time and frequency domain images of the LFMJ.}
	\label{fig6}
\end{figure}
\begin{figure}[!t]
	\vspace{-1em}
	\centering
	\includegraphics[width=3.4in]{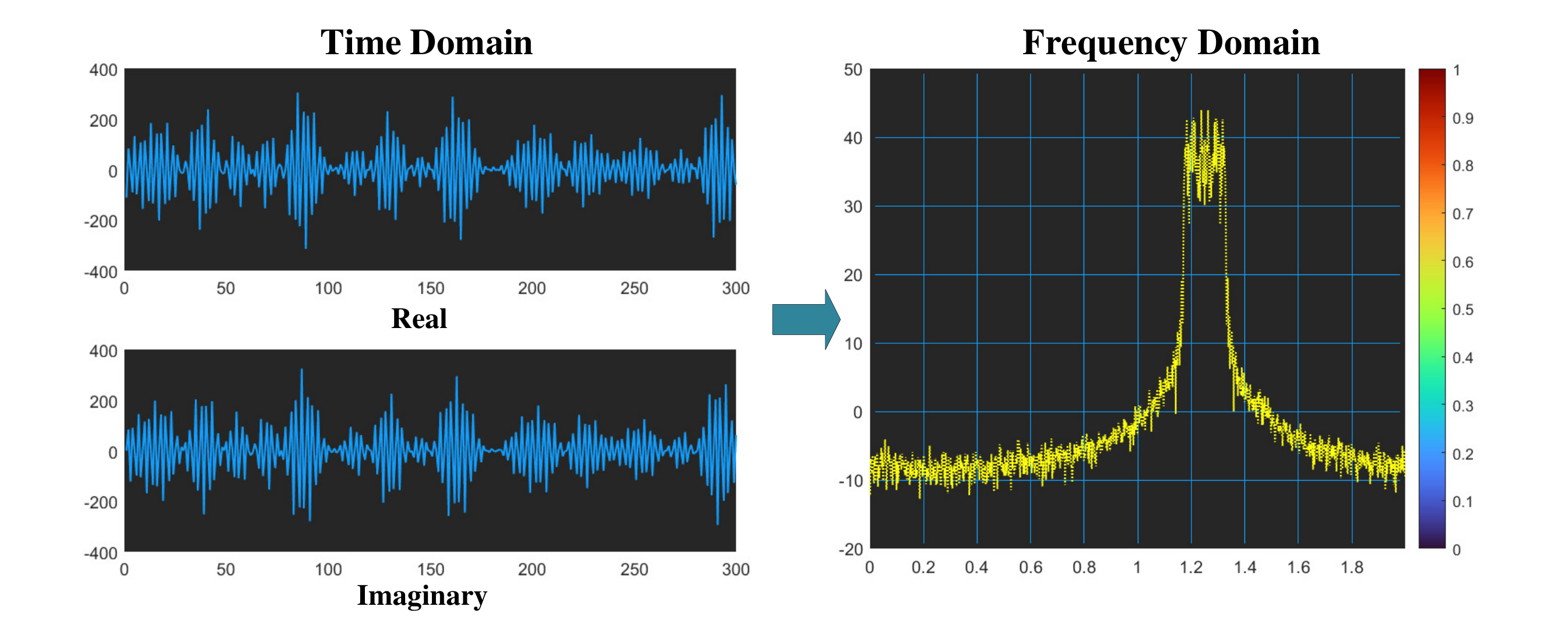}
	\caption{Time and frequency domain images of the PPNJ.}
	\vspace{-1em}
	\label{fig7}
\end{figure}
Periodic Pulse Noise Jamming (PPNJ) is obtained by amplitude modulation of the periodic pulse by partial band noise jamming. The mathematical expression is a piecewise function:
\begin{equation}
	\begin{aligned}
		 J(t)&=U_n(t) U_{T, \tau}(t) \\
		& =\left\{\begin{array}{ccc}
			U_n(t), & k T \leq t \leq k T+\tau, & k \in Z \\
			0, & \text { otherwise }
		\end{array},\right. \\
	\end{aligned}
\end{equation}
where $U_{T,\tau}(t)$ stands for the periodic pulse signal, and the expression is as follows:
\begin{equation}
	U_{T, \tau}(t)=\left\{\begin{array}{ccc}
		1, & k T \leq t \leq k T+\tau, & k \in Z \\
		0, & \text { otherwise }
	\end{array}.\right.
\end{equation}

\par {In this study, we simulate various jamming types, as previously mentioned, to verify the performance of different algorithms. 
	 These jamming signals exhibit 1D signal characteristics in their statistical properties, and they possess significant similarities in terms of attributes such as frequency points and amplitude within their respective domains.
	 
	}

\section{Methodology}

In this section, we propose the CGAN-based fusion 1D-CNN (CGAN-1DCNN) framework for few-shot jamming signal recognition and provide the corresponding algorithm details and processes.
\subsection{The Framework of CGAN-1DCNN Algorithm}
\begin{figure*}[!t]
	\centering
	\includegraphics[width=6.4in]{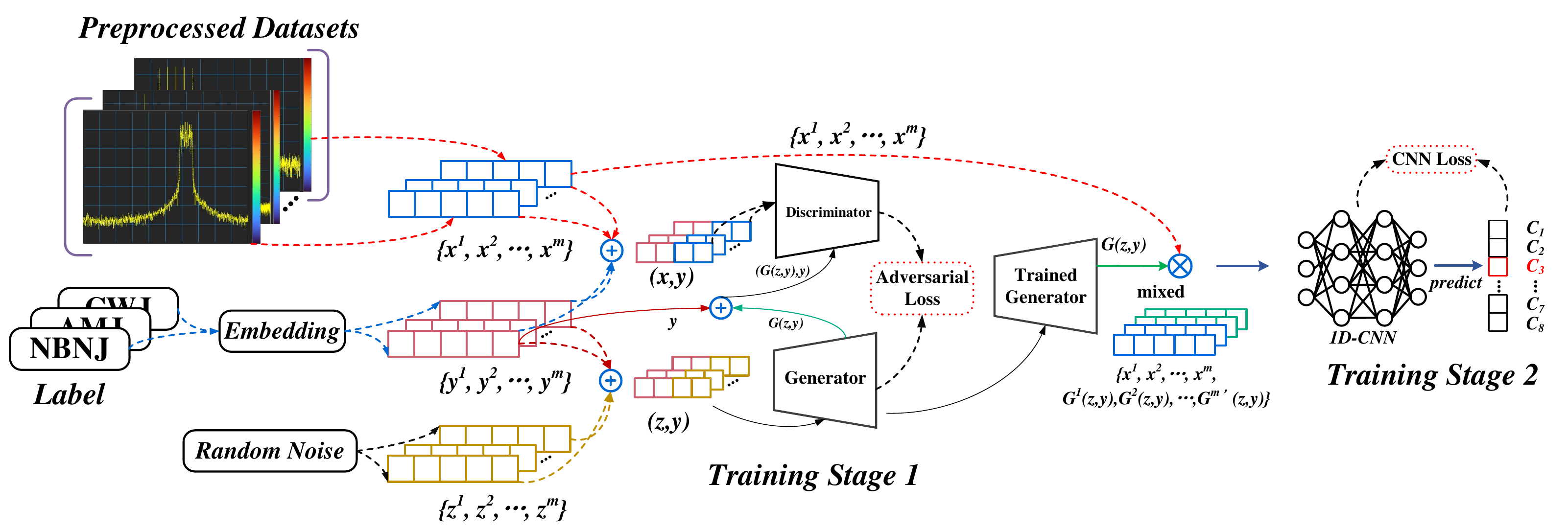}
	\caption{{The overall procedure of CGAN-based fusion CNN framework. The ‘‘+’’ refers to the concatenation of two vectors, and the ‘‘×’’ refers to the mixing of different samples.}}
	\label{fig8}
\end{figure*}
\begin{figure*}[!t]
	\centering
	\includegraphics[width=6.4in]{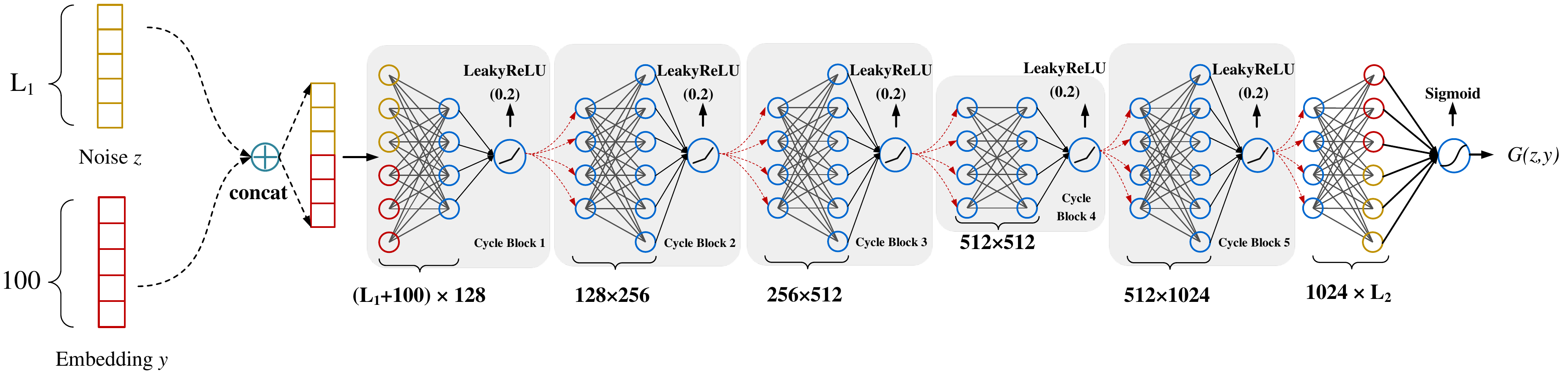}
	\caption{The network architecture of the generator. }
	\label{fig9}
\end{figure*}
%
%
%
\par {This part primarily comprises three key components: i) signal data preprocessing, ii) CGAN training, and iii) 1D-CNN training. 

	The preprocessing phase involves calculating the power spectral density (PSD) of the signal in the frequency domain to produce a 1D real vector, denoted as $x$, with a length of $L_1$. The calculation method is as follows:
\begin{equation}
	x=10*\log(|FFT(I+jQ)|),
\end{equation}
where $I$ and $Q$ are the real and imaginary parts of the signal.

CGAN \cite{2014Conditional} represents an enhanced architecture that introduces conditional encoding information to the foundational GAN structure. It consists of a generator, denoted as $G$, and a discriminator, denoted as $D$. In this paper, we utilize class labels from various jamming signals as the conditioning factor. Through an embedding operation, we generate a 1D vector with a length of 100, referred to as $y$ in Fig. 8. Additionally, $z$ signifies a 1D random noise sequence, obtainable by sampling Gaussian white noise.

To proceed, we concatenate the vectors $x$ and $y$, as well as the vectors $y$ and $z$, and subsequently feed them as inputs into the CGAN module. Then save the $G's$ output $G(z, y)$ upon convergence of training and blend it with the original input $x$ for data augmentation. At last, we feed the combined data into a 1D-CNN classifier for a secondary round of training. Until training convergence, the model outputs the predicted category.
}
\begin{figure*}[!t]
	\centering
	\includegraphics[width=6.4in]{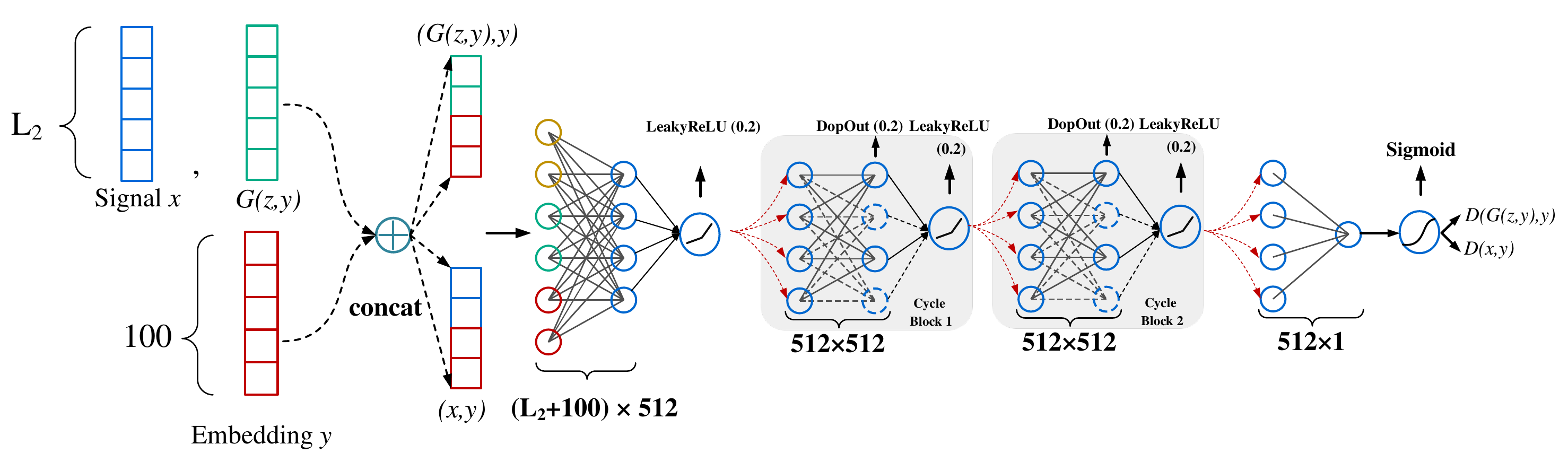}
	\caption{The network architecture of the discriminator.}
	\label{fig10}
\end{figure*}
\begin{figure*}[!t]
	\centering
	\includegraphics[width=6.4in]{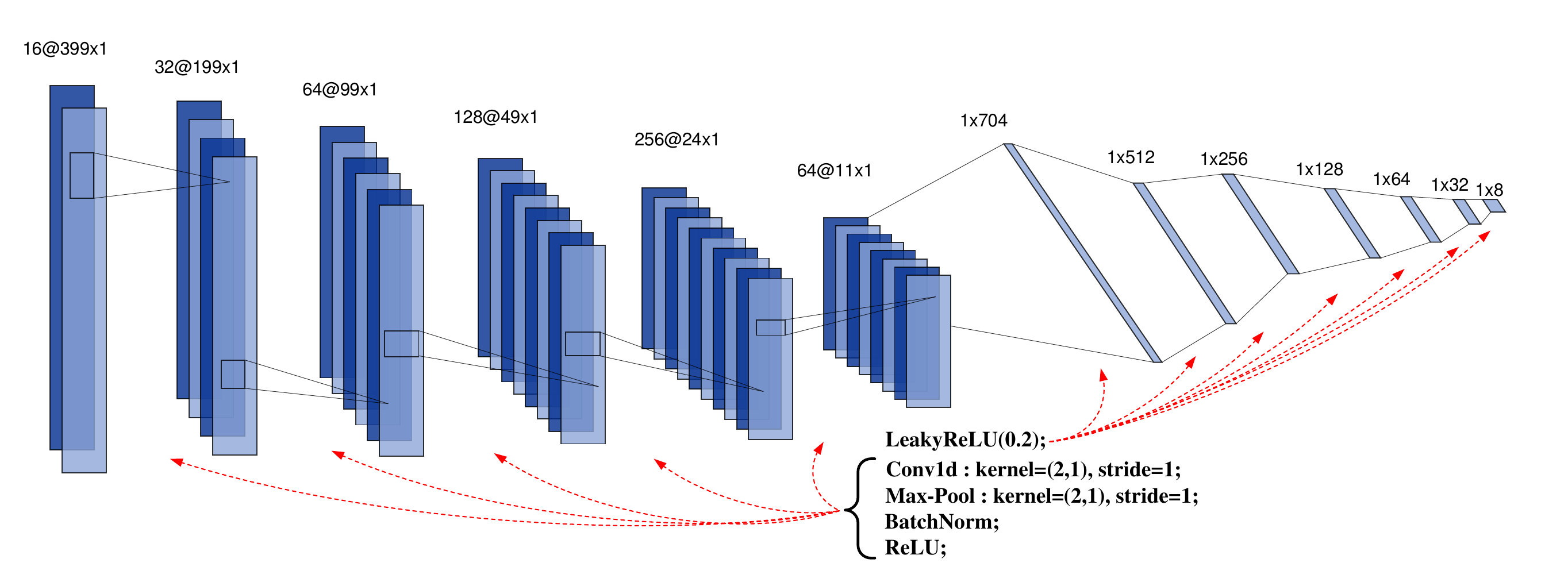}
	\caption{The network architecture of the 1D-CNN.}
	\label{fig10}
\end{figure*}
\subsection{{Network Structure}}
\par {The architectural composition of $G$ entails multiple fully connected layers, as visually exemplified in Fig. 9. $G$ encompasses 5 consecutive cycle blocks, all characterized by an identical structure, comprising two fully connected layers and employing the LeakyReLU activation function. The representation of the number of neurons in connected layers adjacent to each other within distinct blocks is denoted as $(M_1\times M_2)$. $G$ receives the concatenation of vector $y$ and a random noise vector $z$, which is equivalently a 1D vector with an input dimensionality of $(L_1+100)$. Subsequently, $G$ produces a 1D output vector $G(z, y)$ with a length of $L_2$.

The architecture of $D$ closely mirrors that of $G$, with detailed structural components and parameter settings illustrated in Fig. 10. Within this architectural depiction, a dropout layer has been incorporated into the cycle block to mitigate the risk of overfitting. The output denoted as $G(z, y)$ for $G$, and the preprocessed 1D vector $x$ are concatenated with $y$, respectively. Subsequently, this concatenated information is transformed into a 1D vector with a length of $(L_2+100)$. This resulting vector is then fed into $D$, ultimately yielding an output within the range of 0 to 1 through the utilization of the sigmoid function.

From Figs. 9 and 10, both the $G$ and $D$ have an input vector with variable parameters. It can be seen that the input of the $G$ contains a noise sequence with the length of $L_1$, while the input of the $D$ includes a vector with the length of $L_2$. Therefore, we can control the length of the $G's$ output by changing the value of $L_2$ to ensure that it meets the expected signal length. 
For example, to both comprehensively display spectral information and reduce training complexity, we selected a moderate frequency domain sequence length. So, in our experiment, the length of 1D vector $x$ obtained through preprocessing is 800 and then we set $L_2$ to 800. After training, the output $G (z, y)$ of the $G$ has the same length as the original data $x$ and is then fed into the $D$ for $D's$ training.

Fig. 11 shows the architecture of 1D-CNN, which mainly consists of 1D convolution, 1D pooling operation, 1D batch normalization operation, and different activation functions. The red arrows in Fig. 11 indicate the specific operations and parameters between adjacent layers.
\begin{algorithm}
	\caption{Two training stage.}
	\KwIn{Training set $\left\{x^1, x^2, x^3, \ldots, x^m\right\}$ drawn from $p_{data}(x)$;\\
		Embedded label $\left\{y^1, y^2, y^3, \ldots, y^m\right\}$; Random noise $\left\{z^1, z^2, z^3, \ldots, z^m\right\}$ sampled from standard normal distribution; }
	\KwOut{Output predictions for the jamming category ${C_i}$.}
	\emph{Initialize network parameters $\theta_D$, $\theta_G$ and $\theta_{CNN}$}
	
	\For{$iter=1$ to $E_1$}
	{
		\emph{Update $\theta_D$ by gradient descent on $L_D$ (16) in one step while fixing $\theta_{G}$}\\
		\emph{Update $\theta_G$ by gradient descent on $L_G$ (15) in one step while fixing $\theta_{D}$}	
	}
	Fix the $\theta_G$ and draw $G(z,y)$ from the $G$. Then feed $m'$ sample pairs $\{G(z,y),y\}$ and $m$ pairs $\{x,y\}$ into 1D-CNN.\\
	\For{$iter=1$ to $E_2$}
	{
		\emph{Update $\theta_{CNN}$ by gradient descent on $L_{CNN}(17)$}\\
	}	
	Feed the testing set to be predicted into 1D-CNN, return $C_i$.
\end{algorithm}
The activation functions contained in Figs. 9, 10, and 11 are as follows:
\begin{equation}
	\sigma(x)=\frac{1}{1+e^{-x}},
\end{equation}

\begin{equation}
	\operatorname{ReLU}(x)=\left\{\begin{array}{cl}
		x & , x>0 \\
		0 & , x \leq 0
	\end{array},\right.
\end{equation}
\begin{equation}
	\operatorname{LeakyReLU}(x)=\left\{\begin{array}{cl}
		x & , x>0 \\
		\alpha x & , x \leq 0
	\end{array}\right..
\end{equation}
}

\section{{Different Objectives For Training}}
\par {In this part, we present the objectives and specific training methods for two training stages in CGAN-1DCNN framework. The algorithm flow is provided in Algorithm 1.
\subsection{{The Objective Function of Generator}}

At the first training stage, $G$ and $D$ are trained simultaneously. We update parameters for $G$ to minimize $log(1-D(G(z, y)))$ and update parameters for $D$ to maximize $logD(x, y)$ as if they are following the min-max game following the function $V(D, G)$:
\begin{equation}
	\begin{aligned}
		& \min _G \max _D V(D, G) \\
		& =\mathbb{E}_{x \sim p_{\text {data}}(x)}[\log D(x,y)]+\mathbb{E}_{z \sim p_z(z)}[\log (1-D(G(z,y)))],
	\end{aligned}
\end{equation}
where $p_{data}(x)$ and $p_z(z)$ represent the probability distribution of $x$ and sampling noise $z$, respectively.

The batch size used during training is denoted as $N$, and the output of $D$ after receiving the concatenated vector of $y$ and $G(z, y)$ is referred to as $D(G(z,y),y)$. $D(G(z,y),y)$ is an $N\times1$ dimensional vector, where each element represents the prediction output, with values ranging between 0 and 1. To alleviate the difficulty of training, we introduce a 1D vector of length $N$, denoted as $\upsilon_1$, composed entirely of elements equal to 1, and then calculate the Binary Cross Entropy (BCE) loss between $D(G(z, y),y)$ and $\upsilon_1$. The loss function is as follows:
\begin{equation}
	\setlength{\abovedisplayskip}{3pt}
	\begin{aligned}
		&\min _G V(D,G)=L_{G}=L_{BCE}(D(G(z,y),y),\upsilon_{_1}) \\
		&=-\frac{1}{N}\sum_{i=1}^{N}[\upsilon_{1}^{i}\log(D(G(z,y),y)^{i})\\
		&+(1-\upsilon_{1}^{i})\log(1-D(G(z,y),y)^{i})],
	\end{aligned}
\end{equation}
where $i$ represents the subscript of the vector's elements.
\subsection{{The Objective Function of Discriminator}}
\setlength{\belowcaptionskip}{2cm}
The loss function of $D$ consists of two parts. When D receives $x$ and $y$, we calculate the BCE loss of $D(x, y)$ and $\upsilon_1$, denoted as $L_{real}$. When $D$ receives $G(x, y)$ and $y$, we calculate the BCE loss of $D(G(x,y))$ and $\upsilon_0$, denoted as $L_{fake}$. $\upsilon_1$ represents an $N\times1$ dimensional vector with all elements equal to 1 which is similar to $\upsilon_0$. The loss function of D is as follows:
\begin{equation}
	\begin{aligned}
		&\max _D V(D, G)=L_D={L_{real}+L_{fake}} \\ 
		&={L_{BCE}(D(x,y),\upsilon_{1})+L_{BCE}(D(G(z,y),y),\upsilon_{0})}\\
		&={-\frac{1}{N}\sum_{i=1}^N[\upsilon_{1}^i\log(D(x,y)^i)+(1-\upsilon_{1}^i)\log(1-D(x,y)^i)}+\\
		&(\upsilon_{0}^{i}\log(D(G(z,y),y)^i))+(1-\upsilon_{0}^i)\log (1-D(G(z,y),y)^i)].
	\end{aligned}
\end{equation}

It is noteworthy that the training of $G$ and $D$ is not mutually independent. The updates to the network parameters of $G$ and $D$ rely on alternating optimization using the Adaptive Moment Estimation (ADAM)\cite{2014Adam} function on $L_G$ and $L_D$. Thus, convergence during training can be assessed by monitoring the values of $L_G$ and $L_D$.
In general, it is advisable to terminate training when both $L_G$ and $L_D$ exhibit convergence within the range of $0.4$ to $0.6$, which is close to 0.5. When both $L_D$ and $L_G$ are stable around 0.5, it indicates that the game between $D$ and $G$ has reached an equilibrium.

\subsection{{The Objective Function of 1D-CNN}}
The second stage of training primarily focuses on updating the network parameters of the 1D-CNN. After the convergence of the first stage training, the output $G(z, y)$ of $G,$ mixed with the preprocessed 1D real-valued vector $x$ are input into the 1D-CNN. Once training convergence is achieved, the system proceeds to the testing phase for the recognition and classification of the jamming signals to be predicted. The loss function of 1D-CNN is as follows:
\begin{equation}
	L_{CNN}=\frac{1}{N} \sum_i L_i=-\frac{1}{N} \sum_i \sum_{c=1}^M y_{i c} \log \left(p_{i c}\right),
\end{equation}
where $M$ is the number of categories to be predicted, $y_{ic}$ and $p_{ic}$ represent the true category of sample $i$ and the predicted probability that sample $i$ belongs to category $c$, respectively. And the time complexity of Eq. (17) is $O(MN)$.
\begin{table}[]
	\begin{center}
		\caption{Simulation Parameters.}
		\smallskip
		\smallskip
		\label{tab:table1}
		\renewcommand\arraystretch{1.2}
		\resizebox{0.75\columnwidth}{!}{
			
			\begin{tabular}{c c c}
				\toprule[1.2pt]
				\textbf{Jamming Signal Types} & \textbf{Parameter} & \textbf{Value}\\
				\hline
				$CWJ_A$  & $\sqrt{P_J}$&0.1$\sim$50\\&$f_J$&20MHz\\
				\hline
				$CWJ_W$  & $\sqrt{P_J}$&20\\ &$f_J$&20MHz$\sim$30MHz\\
				\hline
				& $U_0$&10$\sim$110\\ AMJ&$f_m$&22MHz$\sim$28MHz\\
				& $\beta_{AM}$&1\\
				\hline		
				MTJ& $N_T$ &2$\sim$6\\ &$f_T$&22MHz$\sim$28MHz\\			
				\hline
				NBNJ  & RBW&0.02$\sim$1\\ 
				\hline
				LFMJ  & $f_H$&2.024$\sim$3MHz\\&$f_L$& -2$\sim$0MHz\\
				\hline
				NAMJ  & $U_0$&2$\sim$50\\&RBW& 0.3\\
				\hline
				& $T$&8000$\sim$18000\\PPNJ&$\tau$&2000$\sim$3000 \\&RBW& 0.2\\
				\hline
				& $f_J$&25MHz\\Common Parameters& $\theta_i$&[0,2$\pi$)\\&Jamming to Noise Ratio (JNR)&-20$\sim$20dB\\
				\toprule[1.2pt]
			\end{tabular}
		}
	\end{center}
\end{table}
\section{Experiments Design}
In this section, we provide detailed explanations primarily concerning software simulation parameters, hardware platform setup intricacies, and algorithm training parameters.
\subsection{Simulation Description of Software-Based Datasets}

%

\par Datasets used in this paper, which include eight different jamming types, follow the equation shown in Section III-A and simulation conditions adhere to the parameters provided in TABLE II. 

In the software simulation experiment, we generate 500 samples for each jamming signal type, where 100 samples are used for few-shot training, and the remaining 400 for test. In order to demonstrate the ascendancy of the CGAN-1DCNN, different few-shot learning algorithms are used for the comparison experiment under the same dataset. 
\begin{enumerate}[]

	\item $SGAN$: SGAN \cite{LuoLL0Y22} uses the semi-supervised learning with GAN to discriminate the deceptive jamming in a radar system under a few-sample dataset.
	\item $1D-CNN$ $Only$: Use the only 1D-CNN architecture shown in Fig. 11 for performance comparison.
	\item $TL-SPWVD$: \cite{Hou2022RadarJammingCI} applies TL to classify 2D TFIs of radar jamming signal under a few-sample dataset and preprocesses them by using the SPWVD algorithm.
	\item $TL-WVD$: TL is also used\cite{Hou2022RadarJammingCI}, but the preprocessing includes generating 2D TFIs implemented by the WVD algorithm.
	\item $S-1DCNN$: The Siamese 1DCNN network \cite{Shao2020ConvolutionalNN} is proposed to classify the radar jamming signal under limited training samples. The network's inputs are 1D real and imaginary parts of the jamming signal.
	\item $ResNet$: ResNet \cite{He2015DeepRL} is a classic classification algorithm which can effectively extract features from 2D image information.
\end{enumerate}

\begin{figure}[!t]
	\centering
	\includegraphics[width=2.9in]{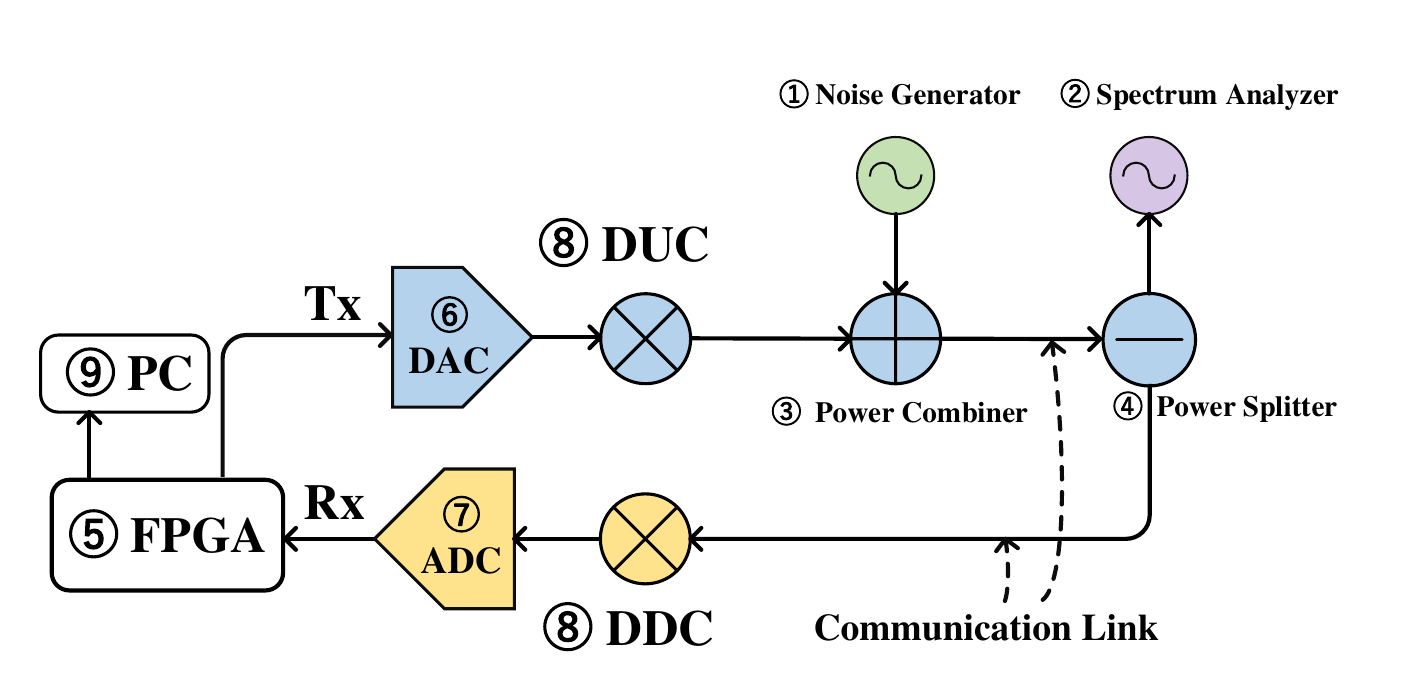}
	\caption{Hardware system model.}
	\label{fig19}
\end{figure}
\begin{figure}[!t]
	\centering
	\includegraphics[width=2.9in]{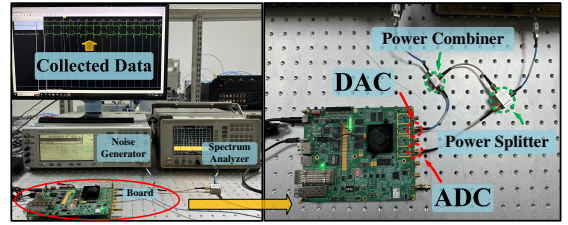}
	\caption{Hardware platform construction details. }
	\label{fig20}
\end{figure}
\begin{table}[!t]
	\begin{center}
		\caption{Hardware Configuration}
		\label{table5}
		\begin{tabular}{c c c} 
			\hline
			\hline
			Number & Function & Hardware model \\
			\hline
			1 &Generate Noise Signal& Agilent E4438C\\
			\hline
			2 &Monitoring/Analyzing & Agilent 8563EC\\
			\hline
			3 &Combine Multiple Signals & ZX10R-14-S+ \\
			\hline
			4 &Split Signals &ZFRSC-183-S+\\
			\hline
			5&Generate Jamming Signal&XCZU27DR\\
			\hline
			6&Convert Digital to Analog&{DAC}\\
			\hline
			7&Convert Analog to Digital&ADC\\
			\hline
			8&Frequency Mixing&RF-LINK\\
			\hline
			9&Data Collecting&PC\\
			\hline
			\hline
		\end{tabular}
	\end{center}
\end{table}

\subsection{Simulation of Hardware-Based Communication Scenario}
\par To further evaluate the effectiveness of our proposed algorithm, we simulate a communication scenario on a hardware platform to acquire a real-world dataset. Herein, we transmit the same signals shown in TABLE II with an intermediate frequency (IF) of 500 MHz and a sampling rate of 163.84 MHz. 
\par The hardware system model conforms to the graphical representation depicted in Fig. 12. Here label 1 denotes a noise generator with the aim of producing noise signals during communication. Label 2 designates a spectrum detector used to monitor and analyze the temporal evolution of the spectrum. Tags 3 and 4 correspond to signal combining and shunting, respectively. The board's main constituents comprise Tags 5, 6, 7, and 8. The Field-Programmable Gate Array (FPGA) is utilized to generate jamming signal sources, while the Digital-to-Analog Converter (DAC) and ADC are employed to effectuate signal conversion. Finally, 
the Personal Computer (PC) serves to collect received data. Table III presents the functionalities and model numbers of the hardware utilized, and Fig. 13 provides a detailed illustration of the constructed hardware platform.



\subsection{Algorithm Parameter Settings}
\par During the training phase, we employ PyTorch as the DL implementation platform for all the experiments, where an RTX 1080 Ti GPU was used to speed up the network training. The TFIs obtained from SPWVD and WVD algorithms are cropped to 224$\times$224$\times$3 size as input for different comparison algorithms. The value of $m'$ is determined to be four times that of the training set size, $m$.
The training epoch $E_1$ and $E_2$ are set to 3200 and 100. $L_1$ is set to 100, and $N$ is set to 128. 
In our experiment, $L_2$ is set to 800, which should be the same length as the preprocessed data $x$. By changing the value of $L_2$, we can generate 1D signal sequences of different lengths.
For the optimizer, we select ADAM with the batch size of 32 and learning rate is selected as 0.0002. 
\renewcommand{\thefigure}{15}
\begin{figure*}[!t]
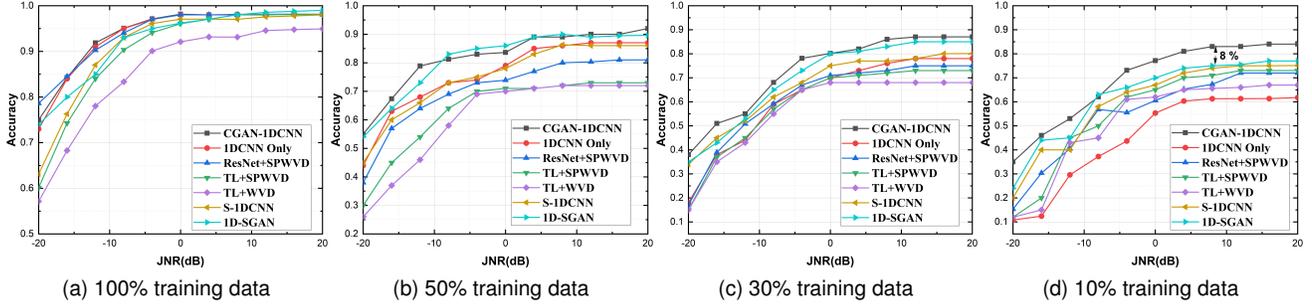

	
	\subfloat[100\% training data]{\includegraphics[width=1.7in]{compare_num_100_revised2.eps}}%
	\subfloat[50\% training data]{\includegraphics[width=1.7in]{compare_num_50_revised2.eps}}%
	\subfloat[30\% training data]{\includegraphics[width=1.7in]{compare_num_30_revised2.eps}}%
	\subfloat[10\% training data]{\includegraphics[width=1.7in]{compare_num_10_revised2.eps}}%
	\caption{Classification accuracy under different capacities of training set.}
\end{figure*}
\renewcommand{\thefigure}{14}
\begin{figure}[!t]
	\centering
	\includegraphics[width=3.4in]{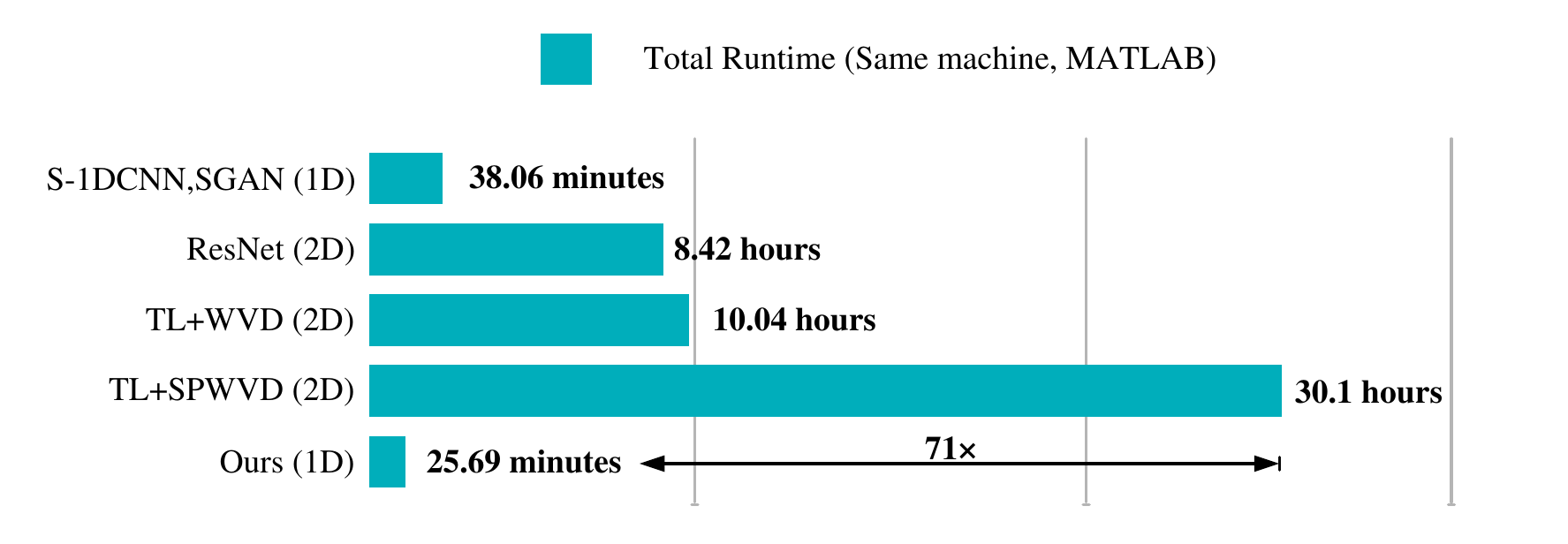}
	\caption{Comparative results on Time Complexity of Different Algorithms.}
	\label{fig13}
\end{figure}

\section{Results and Discussion}
\par
	In this section, we present a comparative analysis of the performance of various algorithms in the context of jamming classification. Additionally, we conduct a performance evaluation and interpretation of the algorithm proposed herein, culminating in the corresponding conclusions.

\subsection{Comparison on Time Complexity Among Different Algorithms}
\par At first, we conduct a comparative analysis of different algorithms, focusing on their preprocessing execution times when applied to generate the same dataset, which contains 500 samples under each class.

Based on the experimental results presented in Fig. 14, our proposed method, which calculates the PSD of jamming signals in the frequency domain as a preprocessing step, minimizes the computational time required to the shortest duration. The method only takes 25.69 minutes, which represents a mere 1/71 of the time consumed by the most time-intensive method. Simultaneously, it is evident from the results that the 2D preprocessing method is more intricate and time-consuming, making real-time classification prediction challenging to implement. Consequently, it lacks practical applicability in real-world communication scenarios.


Furthermore, we evaluate the computation complexity by computing floating-point operations per second (FLOPs) and parameter size in different comparable algorithms. The FLOPs of a convolutional layer are defined as $O(\sum_{l=1}^{D^{*}}M_l^2\cdot K_l^2\cdot C^{*}_{l-1}\cdot C^{*}_l)$, where $D^*$ is the depth of the network, $M$ is the length of the feature map, $C^{*}$ is the number of channels, and $K_l$ represents the kernel of the $l$-th layer. Similarly, the parameter size is computed as $O(\sum_{l=1}^{D^{*}}K_l^2\cdot C^{*}_{l-1}\cdot C^{*}_l+\sum_{l=1}^{D^{*}}M_l^2\cdot C^{*}_l)$.

The comparative results presented in Table IV reveal that, despite CGAN-1DCNN not having the smallest parameter size, it achieves optimal FLOPs, resulting in the lowest time complexity. Considering the overall preprocessing runtime, CGAN-1DCNN exhibits lower comprehensive time complexity compared to other methods, which better meets the applicability requirements in real-world communication scenarios.

\subsection{Performance under Different Capacities of Training set}
In accordance with the description in Section VI-A, the training set derived from software simulation comprises 100 instances, while the testing set encompasses 400 instances. We randomly sampled subsets containing 100\%, 50\%, 30\%, and 10\% of the training set to conduct an in-depth assessment of the algorithm's efficacy. The experimental configurations remained in strict conformity with the specifications outlined in Section VI-C.
\begin{table}[!t]
	\caption{FLOPs and Parameters of Different Algorithms}
	\centering
	\resizebox{0.6\columnwidth}{!}{
		\begin{tabular}{ccc}
			\toprule[1.2pt]
			\textbf{Structure}&\textbf{FLOPs}&\textbf{Parameters}\\
			\midrule
			CGAN-1DCNN&$0.24e^9$&$4.97e^6$\\
			\midrule
			SGAN&$0.35e^9$&$3.50e^6$\\
			\midrule
			S-1DCNN&$4.51e^{10}$&$4.11e^6$\\
			\midrule
			ResNet&$4.1e^{10}$&$2.56e^7$\\	
			\midrule
			TL&$0.83e^9$&$1.25e^6$\\
			\toprule[1.2pt]
		\end{tabular}	
	}
\end{table}
{The outcomes of the experimental analysis are presented in Fig. 15. At lower JNR levels, the classification accuracy is relatively low, and as the JNR increases, the accuracy also improves, ultimately reaching convergence. The reason for this phenomenon lies in the fact that when JNR$<$0, $\frac{P_J}{P_N}$ is also less than 1, indicating that $P_J<P_N$, where $P_J$ and $P_N$ represent the power of the jamming signal and the power of the noise signal, respectively. Under these conditions, the jamming signal is submerged by the noise signal, and the amplitude of the noise becomes the dominant characteristic. Consequently, different types of jamming signals exhibit uniform characteristics resembling those of the noise signal, making accurate classification challenging.

 	 Additionally, when the training subset constitutes 100\% of the overall training set, various algorithms have equivalent performance. 
As the proportion of the training subset diminishes, there is a marked deterioration in the performance of previous algorithms. However, even under the constraints of limited training samples, our algorithm still manages to achieve a higher level of accuracy when compared to other algorithms. Notably, when the training subset represents 10\% of the overall training set, CGAN-1DCNN attains a remarkable 8\% enhancement in accuracy relative to 1D-SGAN, thereby achieving a level commensurate with the state of the art.

In situations involving limited sample sizes as shown in Fig. 15(d), the performance distinction between the 1DCNN Only and CGAN-1DCNN can be attributed to our utilization of CGAN for augmenting the diversity of the training samples. This augmentation effectively aligns the data distribution in the training set more closely with that of the testing set. Consequently, it enhances the neural network's learning and generalization capabilities to a certain extent, resulting in superior recognition accuracy, thus manifesting as a performance improvement.
}}

\subsection{Quality Evaluation of Generated Samples}
\begin{figure}[!t]
	\centering
	\includegraphics[width=3.1in]{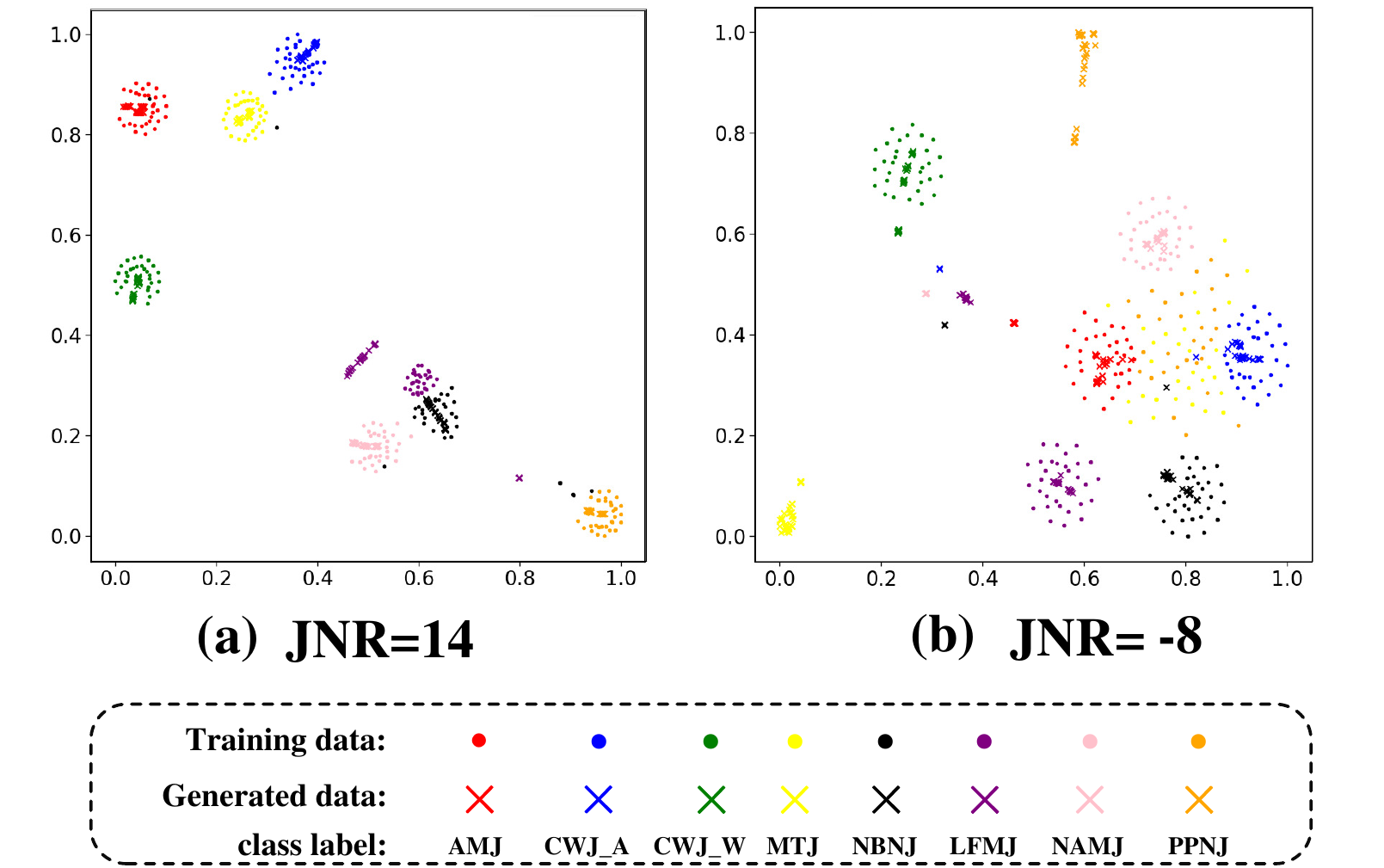}
	\caption{T-SNE dimension reduction result of generator's output data under different JNRs. }
	\label{fig17}
\end{figure}
{\par During the training process, we draw the output $G(z,y)$ of the generator within the CGAN-1DCNN framework and mix it with the original sample $x$ before feeding them into the 1D-CNN module. The similarity in data distribution between $G(z, y)$ and $x$ determines the performance of the algorithm, and now we analyze the samples synthesized by the generator through visual results analysis.

\par The t-SNE\cite{Maaten2008VisualizingDU} results are evaluated through a practical algorithm for reducing data dimensionality. Using the t-SNE algorithm, the higher dimensional features are reduced to two dimensions, which is better visualized on a 2D plane. {The elements in Fig. 16 consist of the $x$, denoted as a point set, and the output data $G(z,y)$, denoted as a cross set.} The Euclidean distance between two elements in space is used to measure their degree of similarity. Precisely, based on the t-SNE results presented in Fig. 16 for two different JNRs, the same type of data elements form a contiguous cluster, indicating a high degree of overlap between the elements.}

\par {In other words, within the same category of jamming signals, the synthesized data $G(z, y)$ generated by $G$ closely approximates the data distribution of $x$. Consequently, mixing $G(z,y)$ with $x$ and then feeding them into the 1D-CNN can compensate for the insufficient original samples, which otherwise results in extremely low classification accuracy. Furthermore, we have depicted the data of $x$ and $G(z,y)$ in curve plots, which are respectively presented in Fig. 17(a)-(b).}

\begin{figure}[!t]
	\flushright
	\subfloat[PSD plots for eight jamming signal types.]{\includegraphics[width=2.5in]{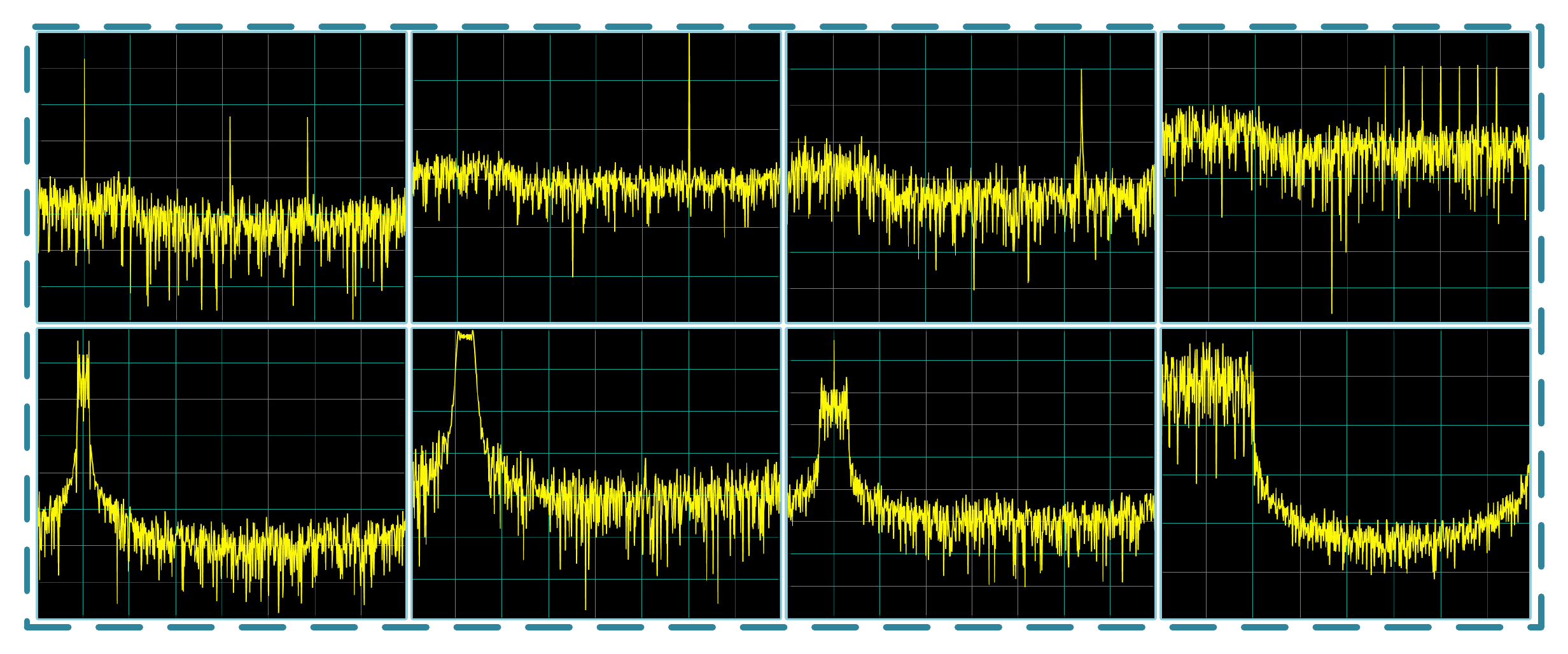}}%
	\hfil
	\newline
	\hfil	
	\subfloat[Synthesized data's PSD plots generated by the generator.]{\includegraphics[width=2.5in]{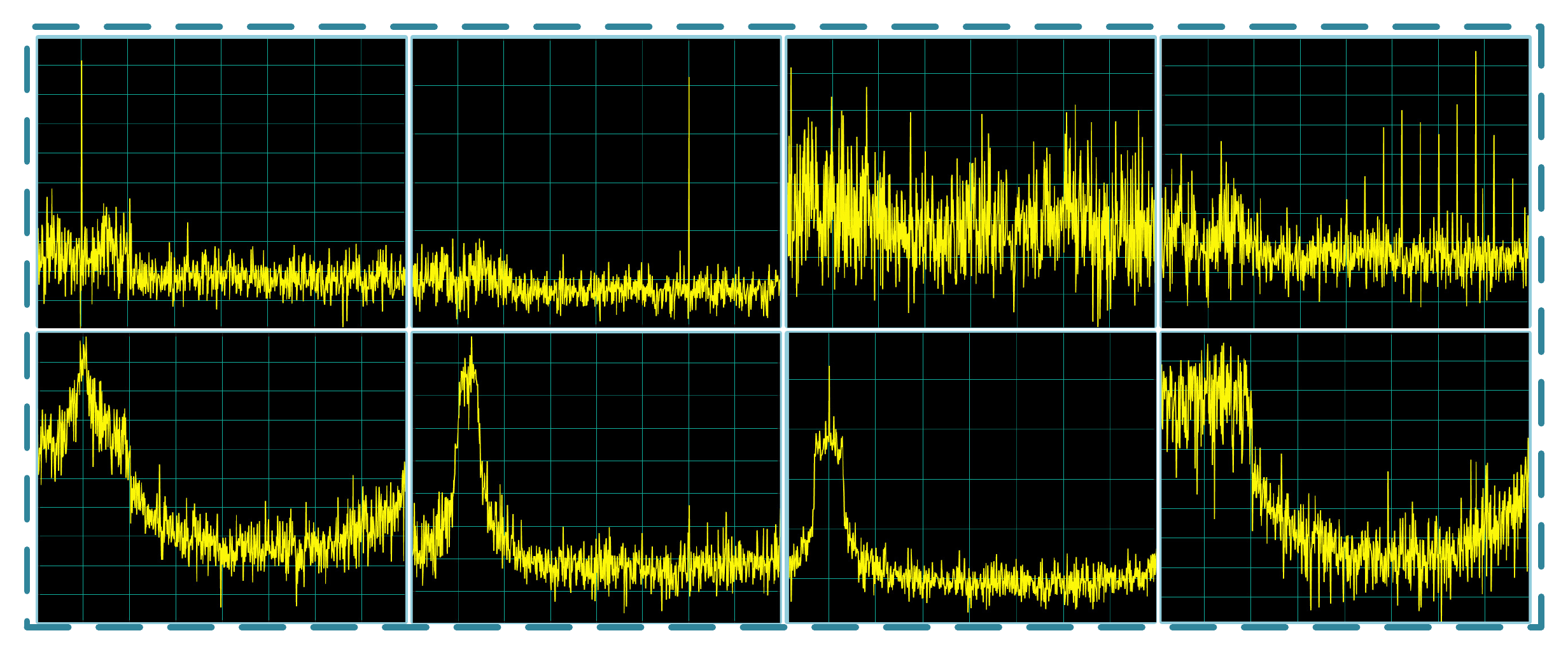}}%
	\hfil
	\newline
	\caption{{Visualization of the synthesized data drawn from the output of CGAN's generator.}}
\end{figure}
\begin{figure}[!t]
\centering
\includegraphics[width=2.6in]{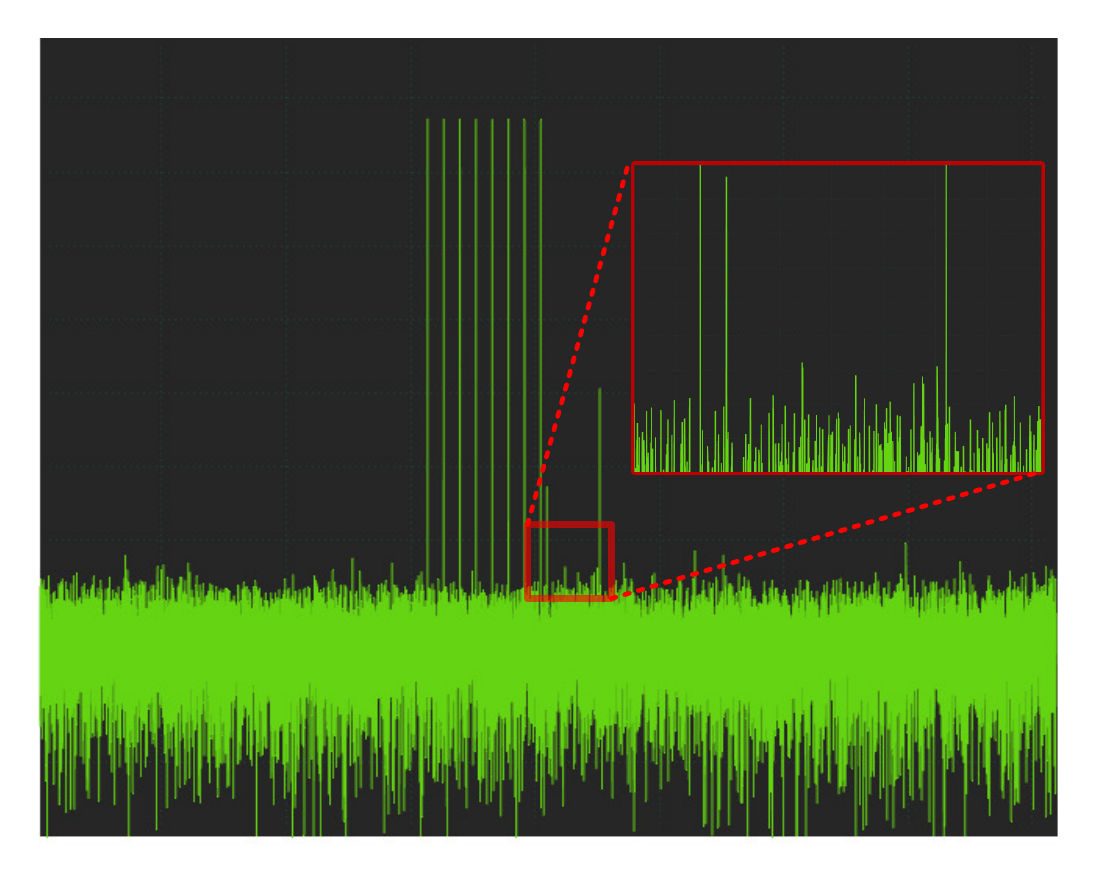}
\caption{{Spurious noise in real-world datasets.}}
\label{fig22}
\end{figure}
\subsection{{Performance of our Algorithm Applied in Real-world Communication Scenario}}

\par {Here, we elaborate the reason to consider the applicability of the algorithms on real-world datasets. Firstly, we need to ensure the practical applicability of the algorithm rather than just staying in the simulation stage. Secondly, it is crucial to acknowledge that there exists a notable disparity in data quality between real-world datasets and software-simulated datasets. The former is derived from analog signals during transmission, subsequently digitized through ADC, and finally acquired by a PC. Throughout this process, a multitude of factors, such as the conversion accuracy of ADC, non-linear characteristics of hardware circuits resulting in harmonics and spurious noise, as well as environmental thermal noise stemming from device heating, collectively influence the quality of the dataset, which further leads to the problem of disordered labeling. The quality of the dataset significantly impacts the algorithm's capacity for feature extraction, thereby influencing its overall performance. Fig. 18 illustrates the occurrence of spurious noise in real-world datasets. Thirdly, the acquisition of an ample quantity of data in real-world scenarios is difficult due to the randomness of jamming. At the same time, we also need to consider whether the algorithm can meet real-time requirements to ensure the identification of jamming types in the shortest possible time.

\begin{figure*}[]
	
	\subfloat[3 samples]{\includegraphics[width=1.7in]{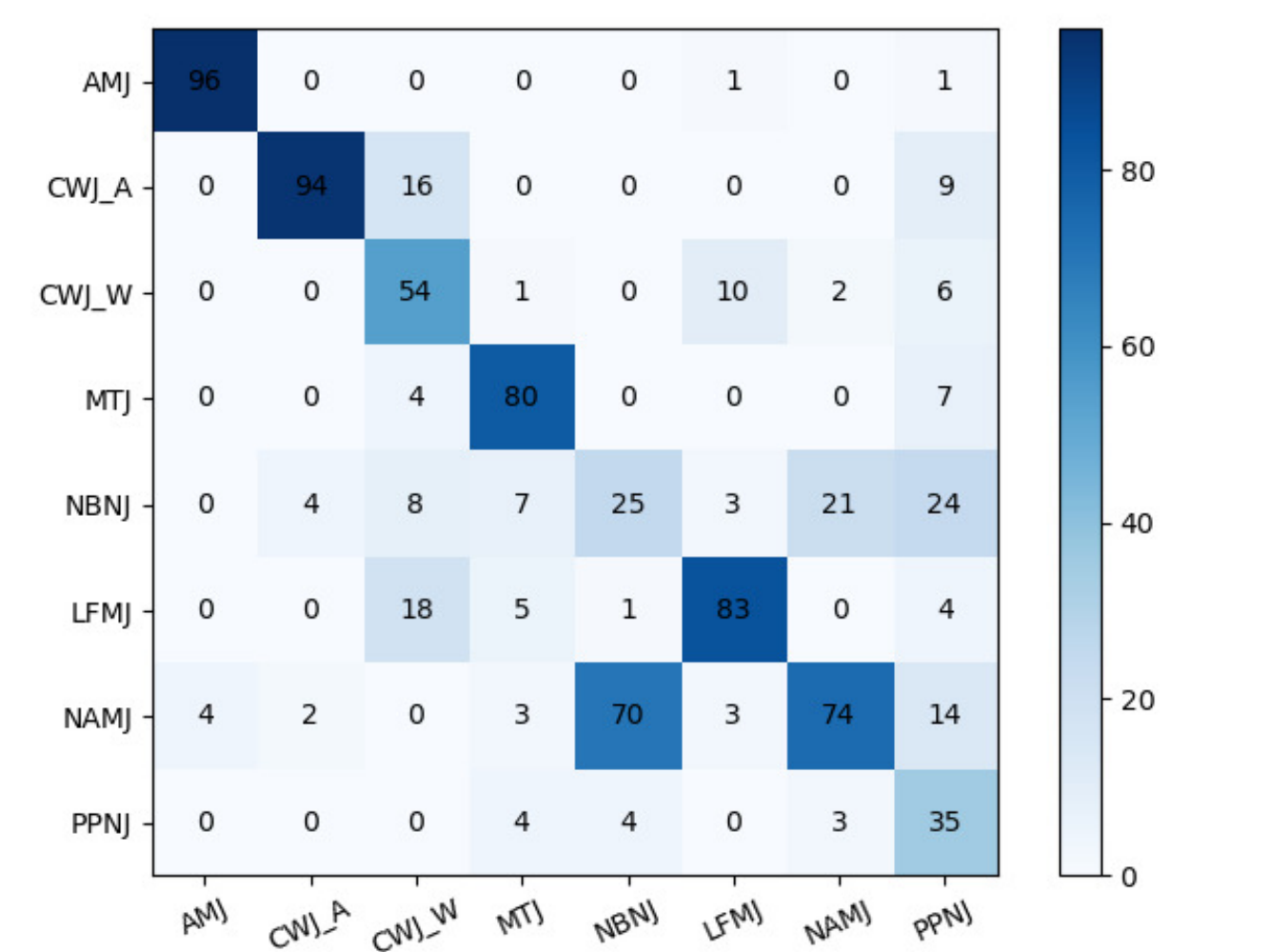}}%
	\subfloat[4 samples]{\includegraphics[width=1.7in]{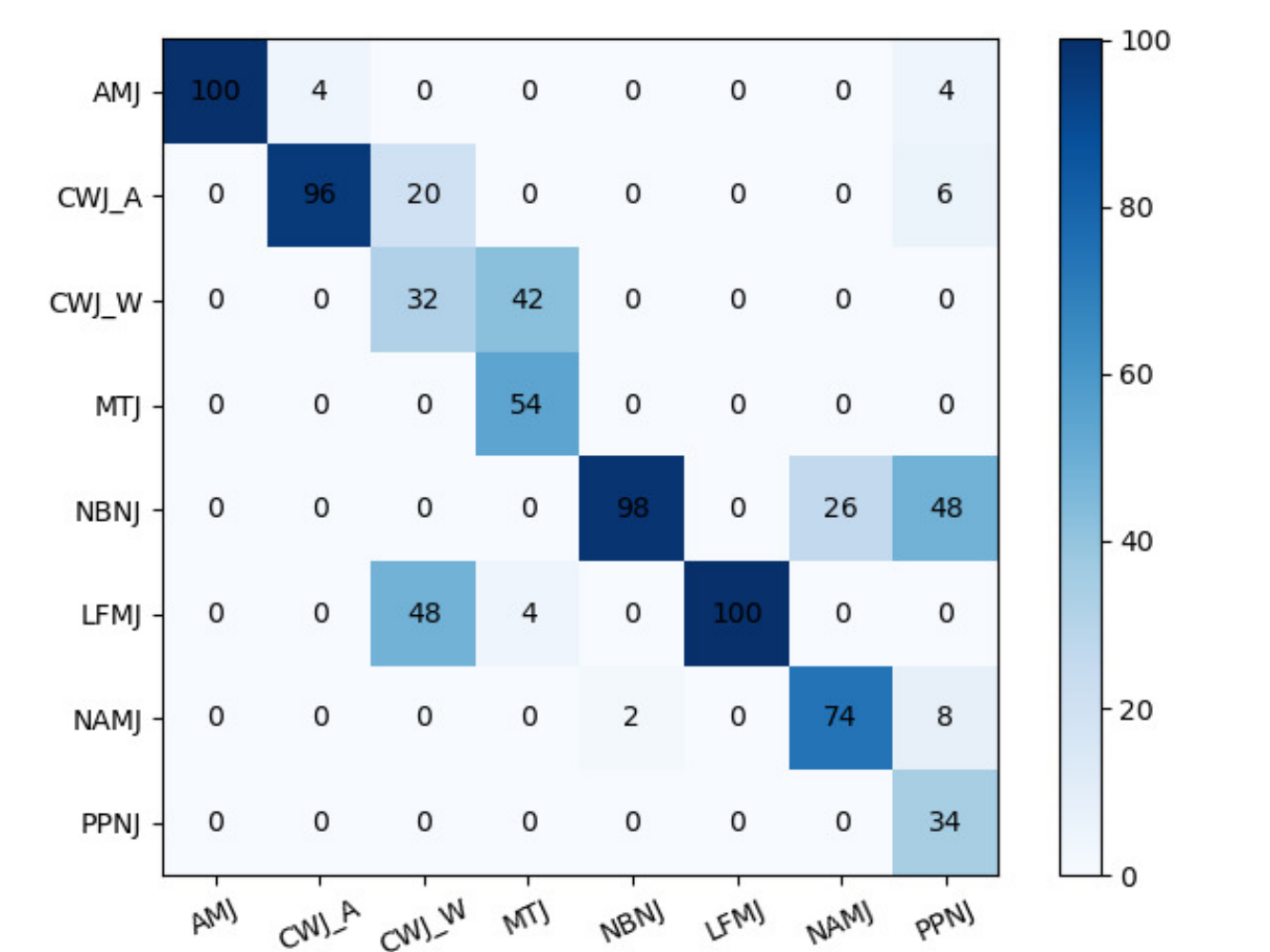}}%
	\subfloat[5 samples]{\includegraphics[width=1.7in]{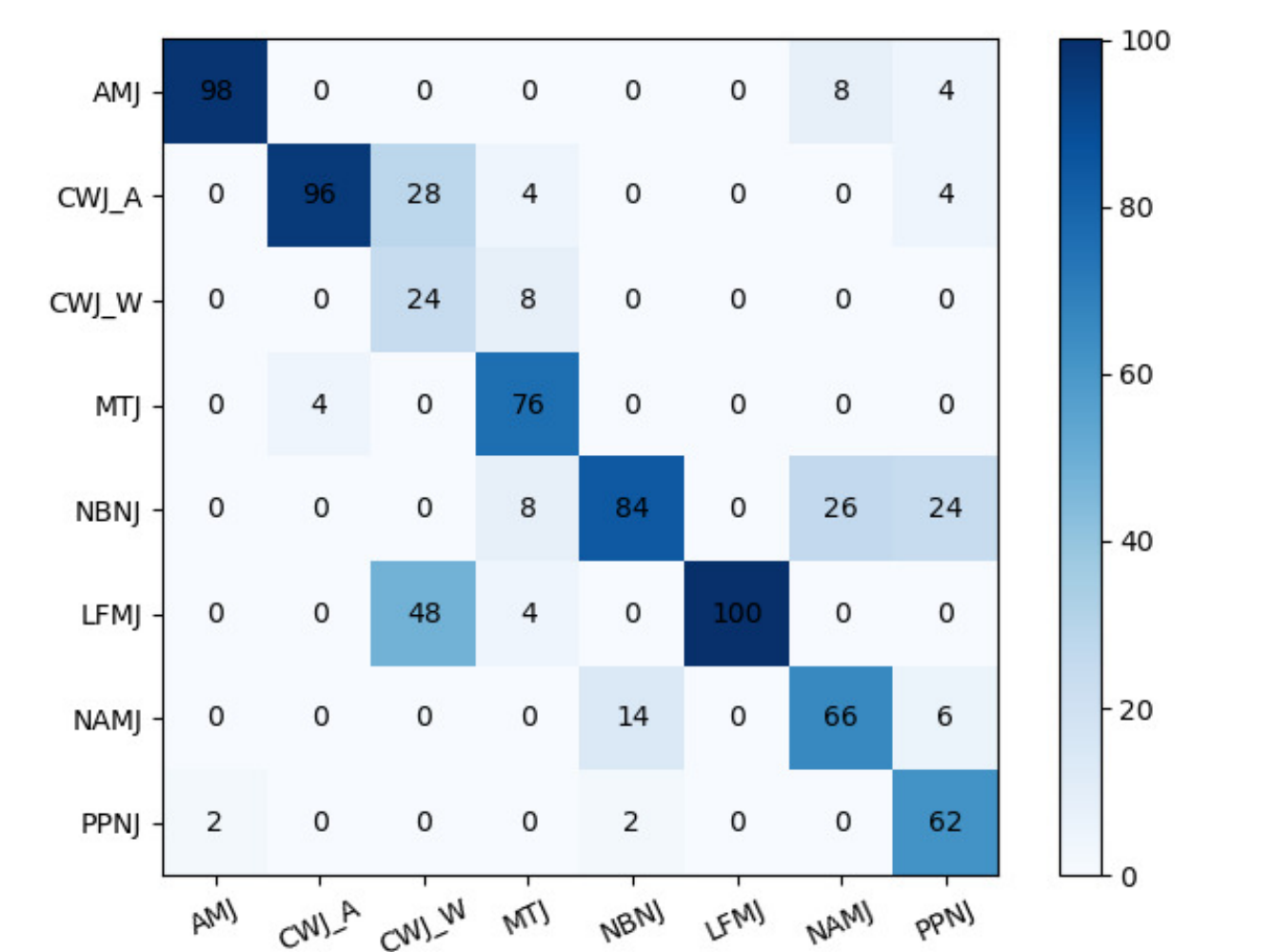}}%
	\subfloat[30 samples]{\includegraphics[width=1.7in]{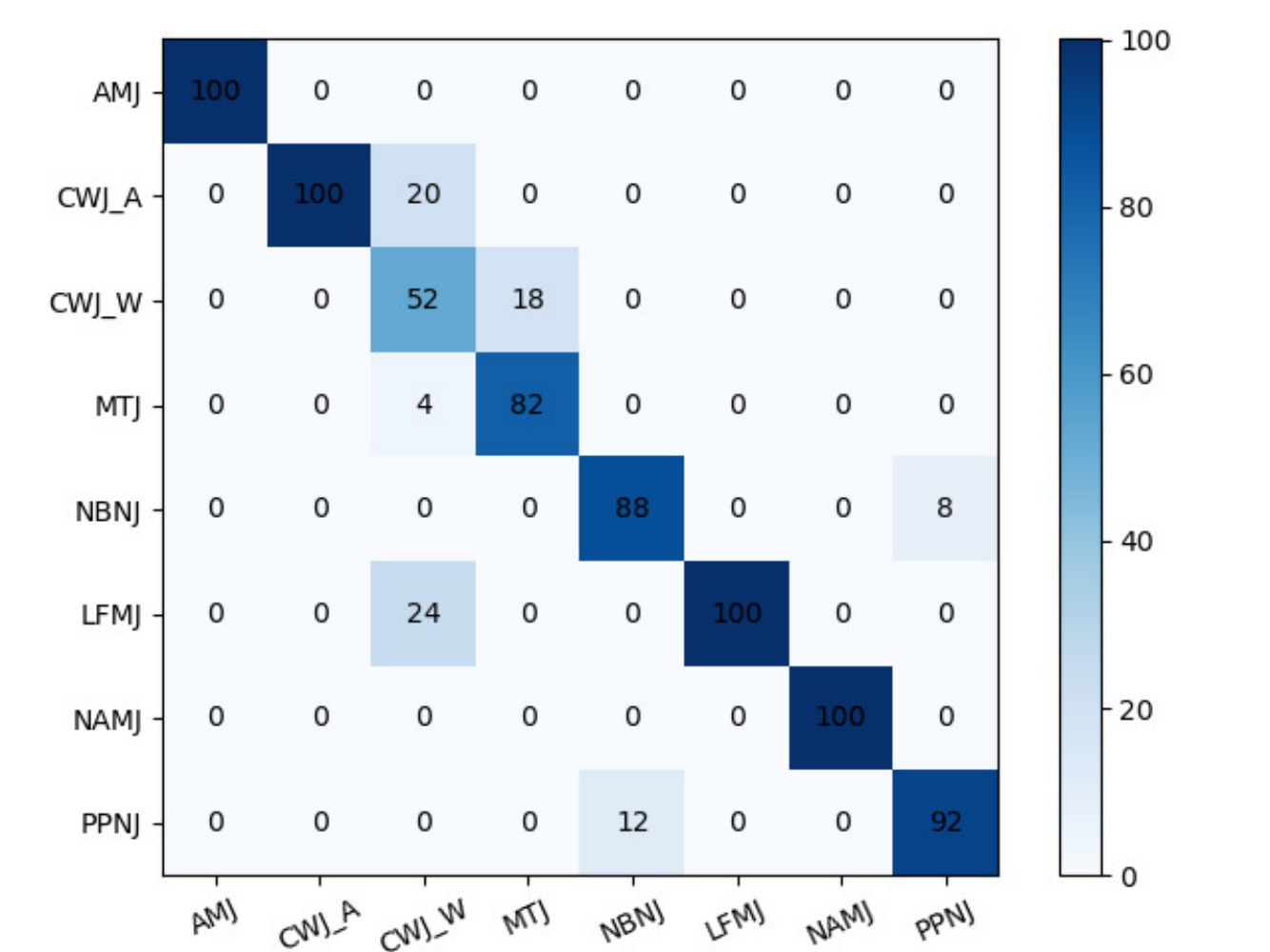}}%
	\caption{Confusion matrix of CGAN-1DCNN on datasets collected by hardware.}
\end{figure*}

\begin{figure}[!t]
	\centering
	\includegraphics[width=2.6in]{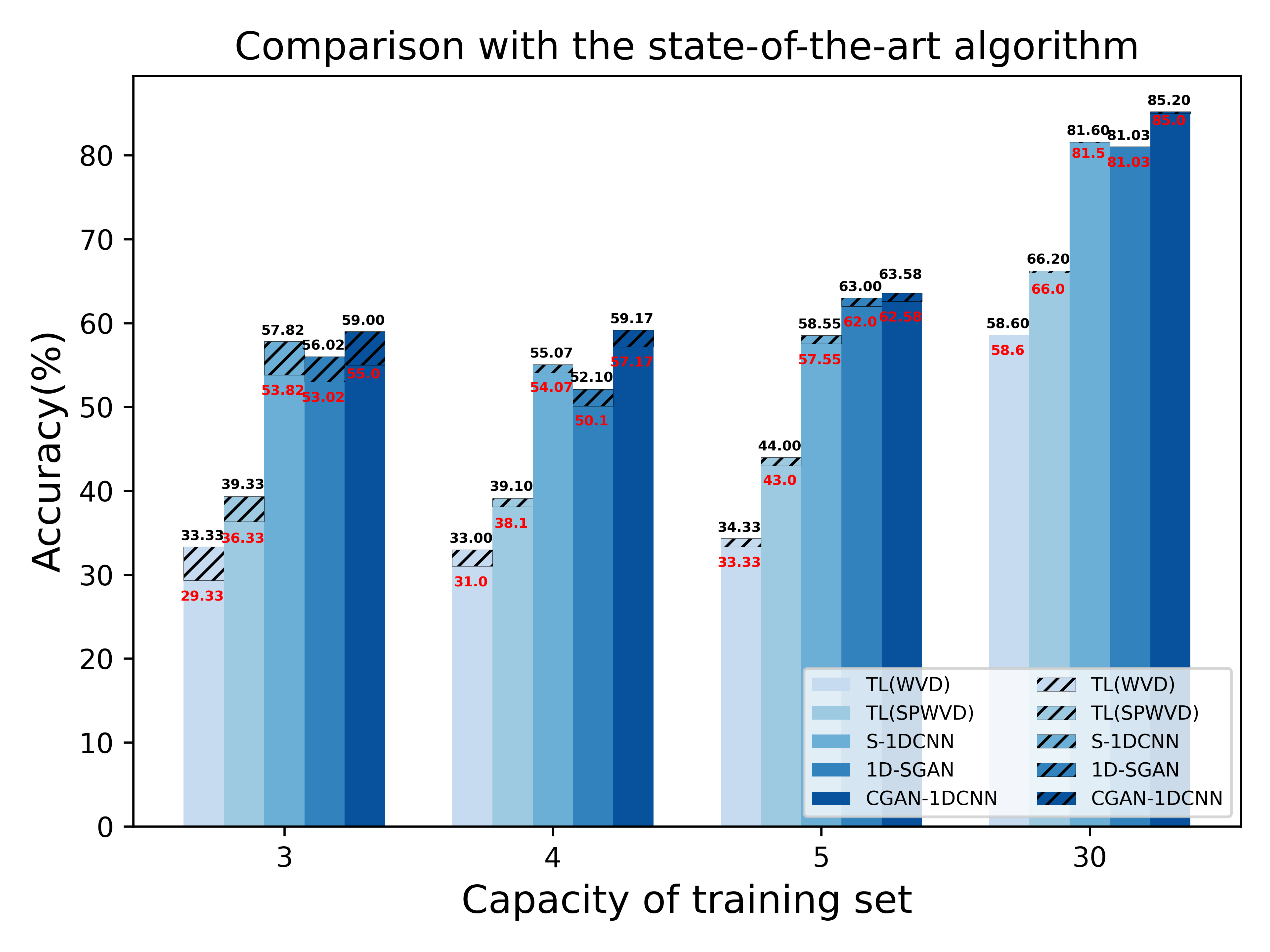}
	\caption{Comparison with the state-of-the-art algorithm under real-world and software-simulated datasets.}
	\label{fig22}
\end{figure}
We employ the hardware system model provided in Section VI-B to transmit eight distinct jamming signals at the TX. We adjusted the JNR within the range of -20 to 20 dB by controlling the power of the noise generator. In the first phase, we collected data to serve as the training set for model training, while in the second phase, we collected data to be used for real-time prediction testing.

During the first phase, we collected 3, 4, 5, and 30 training samples for each category of jamming signals as the training set to conduct offline training following the process in Section V. In the second phase, we randomly selected 100 samples for each category to constitute the testing set, which is waiting for prediction. Fig. 19 presents the confusion matrix representing the prediction outcomes of our algorithm.
}

Based on the analysis of the confusion matrix in Fig. 19, it has been determined that the algorithm is inadequate in classifying two CWJ types, called $CWJ_A$ and $CWJ_W$, characterized by amplitude and frequency changes when only 3, 4, and 5 samples are utilized. This outcome is justifiable as the amount of feature information that can be extracted from such a small sample size is limited, and the recorded signal's data quality is mediocre, making it difficult to detect subtle changes. However, by increasing the number of samples to 30, the algorithm can achieve a higher classification accuracy. 
\par {We have compared the performance with several high-performing algorithms introduced in Section VI-A.}
  Based on the simulation results obtained from our experiments, we observed that our prediction accuracy outperforms their proposed algorithms when using the same datasets. The comparison results are shown in Fig. 20.
  
  In addition, Fig. 20 shows the classification results of the same algorithm applied to two different datasets, where the red font data represents the classification results achieved by the algorithm on real-world datasets, and the black font data represents the classification results achieved on software-simulated datasets. It can be seen that the classification accuracy of the same algorithm for real-world datasets is lower than that for software-simulated datasets. At the same time, when the number of training samples is 3, 4, and 5, the performance difference between two datasets on the same algorithm is significant, until the number of training samples increases to 30, the performance difference gradually decreases. 
  
  This phenomenon can be attributed to the pronounced impact of hardware noise and non-linear factors of hardware on the quality of data when the sample size is extremely small. These elements significantly elevate the bias in data characteristics, consequently affecting the classification outcomes. Such conditions highlight the heightened sensitivity of hardware-dependent datasets to training sample size, in contrast to software-simulated counterparts.
  
  
\section{Conclusions}

{In this paper, we have proposed the CGAN-1DCNN algorithm with a low-time complexity preprocessing step and then conducted algorithm validation on limited software simulation datasets and real-world collected datasets from two aspects: runtime complexity and classification accuracy. The experimental results indicate that our proposed method outperforms existing few-shot learning methods both in time complexity and performance, which has achieved an 8\% improvement in accuracy under a limited software-based training set. Furthermore, we validate its generalization ability using relevant datasets collected by hardware. Comparative results show that our algorithm has achieved extremely high classification accuracy for low-quality and low-precision signal datasets with limited sample sizes. Our research provides new insights for future research on signal processing algorithms and enhances the comprehensiveness of experiments beyond software simulation.}

\bibliographystyle{IEEEtran}
\bibliography{Ref}



\end{document}